# Synchron - An API and Runtime for Embedded Systems

**Abhiroop Sarkar** ✉ ⓘ
Chalmers University, Sweden

**Bo Joel Svensson** ✉ ⓘ
Chalmers University, Sweden

**Mary Sheeran** ✉ ⓘ
Chalmers University, Sweden

## Abstract

Programming embedded systems applications involves writing concurrent, event-driven and timing-aware programs. Traditionally, such programs are written in low-level machine-oriented programming languages like C or Assembly. We present an alternative by introducing Synchron, an API that offers high-level abstractions to the programmer while supporting the low-level infrastructure in an associated runtime system and one-time-effort drivers.

Embedded systems applications exhibit the general characteristics of being (i) concurrent, (ii) I/O–bound and (iii) timing-aware. To address each of these concerns, the Synchron API consists of three components - (1) a Concurrent ML (CML) inspired message-passing concurrency model, (2) a message-passing–based I/O interface that translates between low-level interrupt based and memory-mapped peripherals, and (3) a timing operator, `syncT`, that marries CML's `sync` operator with timing windows inspired from the TinyTimber kernel.

We implement the Synchron API as the bytecode instructions of a virtual machine called SynchronVM. SynchronVM hosts a Caml-inspired functional language as its frontend language, and the backend of the VM supports the STM32F4 and NRF52 microcontrollers, with RAM in the order of hundreds of kilobytes. We illustrate the expressiveness of the Synchron API by showing examples of expressing state machines commonly found in embedded systems. The timing functionality is demonstrated through a music programming exercise. Finally, we provide benchmarks on the response time, jitter rates, memory, and power usage of the SynchronVM.

# 1 Introduction

Embedded systems are ubiquitous. They are pervasively found in application areas like IoT, industrial machinery, cars, robotics, etc. Applications running on embedded systems tend to embody three common characteristics:

1. They are *concurrent* in nature.
2. They are predominantly *I/O–bound* applications.
3. A large subset of such applications are *timing-aware*.

This list, although not exhaustive, captures a prevalent theme among embedded applications. Programming these applications involves interaction with callback-based driver APIs like the following from the Zephyr RTOS[12]:

▪ **Listing 1** A callback-based GPIO driver API

```
int gpio_pin_interrupt_configure(const struct device *port
                                , gpio_pin_t pin
                                , gpio_flags_t flags);
void gpio_init_callback(struct gpio_callback *callback
                       , gpio_callback_handler_t handler
                       , gpio_port_pins_t pin_mask);
int gpio_add_callback(const struct device *port
                     , struct gpio_callback *callback);
```

Programming with such APIs involves expressing complex state machines in C, which often results in difficult-to-maintain and elaborate state-transition tables. Moreover, C programmers use error-prone shared-memory primitives like *semaphores* and *locks* to mediate interactions that occur between the callback-based driver handlers.

In modern microcontroller runtimes, like MicroPython [13] and Espruino (Javascript) [39], higher-order functions are used to handle callback-based APIs:

▪ **Listing 2** Driver interactions using Micropython

```
def callback(x):
#...callback body with nested callbacks...

extint = pyb.ExtInt(pin, pyb.ExtInt.IRQ_FALLING
                   , pyb.Pin.PULL_UP, callback)
ExtInt.enable()
```

The function `callback(x)` from Line 1 can in turn define a callback action `callback2`, which can further define other callbacks leading to a cascade of nested callbacks. This leads to a form of *accidental* complexity, colloquially termed as *callback-hell* [22].

We present Synchron, an API that attempts to address the concerns about callback-hell and shared-memory concurrency while targeting the three characteristics of embedded programs mentioned earlier by a combination of:

1. A message-passing–based concurrency model inspired from Concurrent ML.
2. A message-passing–based I/O interface that unifies concurrency and I/O.
3. A notion of time that fits the message-passing concurrency model.

Concurrent ML (CML) [27] builds upon the synchronous message-passing–based concurrency model CSP [17] but adds the feature of composable first-class *events*. These first-class events allow the programmer to tailor new concurrency abstractions and express application-specific protocols. Additionally, a synchronous concurrency model renders linear control-flow to a program, as opposed to bottom-up, non-linear control flow exhibited by asynchronous callback-based APIs.

Synchron extends CML's message-passing API for software processes to I/O and hardware interactions by modelling the external world as a process through the `spawnExternal` operator. As a result, the standard message-passing functions such as `send, receive` etc. are applicable for handling I/O interactions, such as asynchronous driver interrupts. The overall API design allows efficient scheduling and limited power usage of programs via an associated runtime.

For timing, Synchron has the `syncT` operator, drawing inspiration from the TinyTimber kernel [21], that allows the specification of baseline and deadline windows for invocation of a method in a class. In TinyTimber, `WITHIN( B, D, &obj, meth, 123 );` expresses the desire that method `meth` should be run at the earliest at time `B` and finish within a duration of `D`. Our adaptation of this API, `syncT B D evt`, takes a baseline, deadline and a CML *event* (evt) as arguments and obeys similar semantics as `WITHIN`.

The Synchron API is implemented in the form of a bytecode-interpreted virtual machine (VM) called SynchronVM. The bytecode instructions of the VM correspond to the various operations of the Synchron API, such that any language hosted on the VM can access Synchron's concurrency, I/O and timing API for embedded systems.

Internally, the SynchronVM runtime manages the scheduling and timing of the various processes, interrupt handling, memory management, and other bookkeeping infrastructure. Notably, the runtime system features a low-level bridge interface that abstracts it from the platform-dependent specifics. The bridge translates low-level hardware interrupts or memory-mapped I/O into software messages, enabling the SynchronVM application process to use the message-passing API for low-level I/O.

**Contributions**

- We identify three characteristic behaviours of embedded applications, namely being (i) concurrent, (ii) I/O–bound, and (iii) timing-aware, and propose a combination of abstractions (the Synchron API) that mesh well with each other and address these requirements. We introduce the API in Section 3.
- **Message-passing–based I/O.** We present a uniform message-passing framework that combines *concurrency* and *callback–based I/O* to a single interface. A software message or a hardware interrupt is identical in our programming interface, providing the programmer with a simpler message-based framework to express concurrent hardware interactions. We show the I/O API in Section 3.2 and describe the core runtime algorithms to support this API in Section 4.
- **Declarative state machines for embedded systems.** Combining CML primitives with our I/O interface allows presenting a declarative framework to express state machines, commonly found in embedded systems. We illustrate examples of representing finite-state machines using the Synchron API in Sections 6.1 and 6.2.
- **Evaluation.** We implement the Synchron API and its associated runtime within a virtual machine, SynchronVM, described in Section 5. We illustrate the practicality and expressivity of our API by presenting three case studies in Section 6, which runs on the STM32 and NRF52 microcontroller boards. Finally, we show response time, memory and power usage, jitter rates, and load testing benchmarks on the SynchronVM in Section 7.

## 2 Motivation

• **Concurrency and IO.** In embedded systems, concurrency takes the form of a combination of callback handlers, interrupt service routines and possibly a threading system, for example threads as provided by ZephyrOS, ChibiOS or FreeRTOS. The callback style of programming is complicated but offers benefits when it comes to energy efficiency. Registering a callback with an Interrupt Service Routine (ISR) allows the processor to go to sleep and conserve power until the interrupt arrives.

An alternate pattern to restore the linear control flow of a program is the event-loop pattern. As the name implies, an event-loop based program involves an infinitely running loop that handles interrupts and dispatches the corresponding interrupt-handlers. An event-loop based program involves some delicate plumbing that connects its various components. Listing 3 shows a tiny snippet of the general pattern.

**Listing 3** Event Loop

```
1 void eventLoop(){
2   while (1) {
```

```
3   switch(GetNextEvent()) {
4      case GPIO1 : GPIO1Handler();
5      case GPIO2 : GPIO2Handler();
6      ....
7      default : goToSleep(); // no events
8  }}}
9
10 GPIO1Handler(){ ... } // must not block
11 GPIO2Handler(){ ... } // must not block
12
13 //when interrupt arrives write to event queue and wake up the while loop
14 GPIO1_IRQ(){....}
15 GPIO2_IRQ(){....}
```

Programs like the above are an improvement over callbacks, as they restore the linear control-flow of a program, which eases reasoning. However, such programs have a number of weaknesses - (i) they enforce constraints on the blocking and non-blocking behaviours of the event handlers, (ii) programmers have to hand-code elaborate plumbings between the interrupt-handlers and the event-queue, (iii) they are highly *inextensible* as extension requires code modifications on all components of the event-loop infrastructure, and (iv) they are instances of clearly concurrent programs that are written in this style due to lack of native concurrency support in the language.

Although there are extensions of C to aid the concurrent behaviour of event-loops, such as protothreads [8] or `FreeRTOS Tasks`, the first three listed problems still persist. The main infinite event loop unnecessarily induces a tight coupling between unrelated code fragments like the two separate handlers for GPIO1 and GPIO2. Additionally, this pattern breaks down the abstraction boundaries between the handlers.

• **Time.** A close relative of concurrent programming for embedded systems is *real-time programming*. Embedded systems applications such as digital sound cards routinely exhibit behaviour where the time of completion of an operation determines the correctness of the program. Real-time programs, while concurrent, differ from standard concurrent programs by allowing the programmer to override the fairness of a *fair* scheduler.

For instance, the `FreeRTOS Task` API allows a programmer to define a static *priority* number, which can override the *fairness* of a task scheduler and customise the emergency of execution of each thread. However, with a limited set of priorities numbers (1 - 5) it is very likely for several concurrent tasks to end up with the same priority, leading the scheduler to order them fairly once again. A common risk with priority-based systems is to run into the *priority inversion problem* [33], which can have fatal consequences on hard real-time scenarios. On the other hand, high-level language platforms for embedded systems such as MicroPython [13] do not provide any language support for timing-aware computations.

**Problem Statement.** We believe there exists a gap for a high-level language that can express concurrent, I/O–bound, and timing-aware programs for programming resource-constrained embedded systems. We outline our key idea to address this gap below.

## 2.1 Key Ideas

Our key idea is the Synchron API, which adopts a synchronous message-passing concurrency model and extends the message-passing functionality to all I/O interactions. Synchron also introduces *baselines* and *deadlines* for the message-passing, which consequently introduces a notion of time into the API. The resultant API is a collection of nine operations that can express (i) concurrency, (ii) I/O, and (iii) timing in a uniform and declarative manner.

**The external world as processes.** The Synchron API models all external drivers as processes that can communicate with the software layer through message-passing. Synchron's `spawnExternal` operator treats an I/O peripheral as a process and a hardware interrupt as a message from the corresponding process. Fig. 1 illustrates the broad idea.

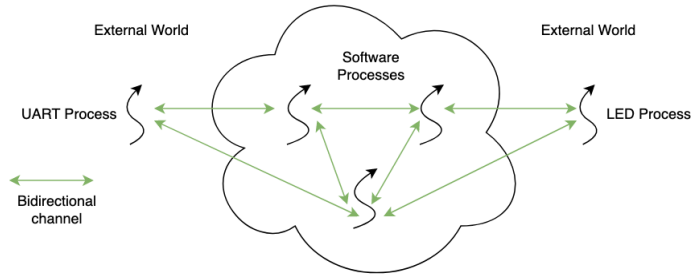

**Figure 1** Software processes and hardware processes interacting

The above design relieves the programmer from writing complex callback handlers to deal with asynchronous interrupts. The synchronous message-passing–based I/O renders a linear control-flow to I/O-bound embedded-system programs, allowing the modelling of state machines in a declarative manner. Additionally, the message-passing framework simplifies the hazards of concurrent programming with shared-memory primitives (like `FreeRTOS semaphores`) and the associated perils of maintaining intricate locking protocols.

**Hardware-Software Bridge.** The Synchron runtime enables the seamless translation between software messages and hardware interrupts. The runtime does hardware interactions through a low-level software *bridge* interface, which is implemented atop the drivers supplied by an OS like Zephyr/ChibiOS. The *bridge* layer serialises all hardware interrupts into the format of a software message, thereby providing a uniform message-passing interaction style for both software and hardware messages. Fig. 2 shows the overall architecture of Synchron.

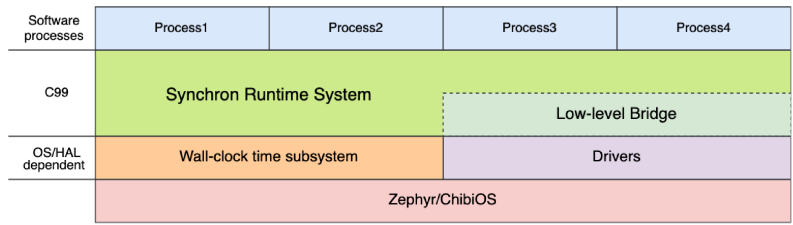

**Figure 2** The Synchron Architecture

**Timing.** The final key component of the Synchron API is the real-time function, `syncT`, that instead of using a static priority for a thread (like Ada, RT-Java, FreeRTOS, etc.), borrows the concept of a dynamic priority specification from TinyTimber [21].

The `syncT` function allows specifying a *timing window* by stating the baseline and deadline of message communication between processes. The logical computational model of Synchron assumes to take zero time and hence the time required for communication determines the timing window of execution of the entire process. As the deadline of a process draws near, the Synchron runtime can choose to dynamically change the priority of a process while it is running. Fig. 3 illustrates the idea of dynamic priority-change.

Fig. 3 shows how a scheduler can choose to prioritise a second process over a running, *timed* process, even though the running process has a deadline in the future. In practice, a scheduler needs to be able to pause and resume processes to support the above, which is possible in the Synchron runtime. The `syncT` function, thus, fits fairly well with the rest of the API and provides a notion of time to the overall programming interface.

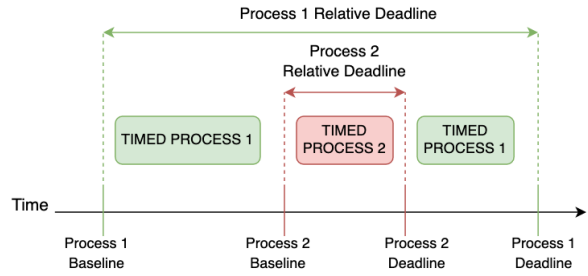

**Figure 3** Dynamic priority change with `syncT`

The combination of `syncT`, `spawnExternal` and the CML-inspired synchronous message-passing concurrency model constitutes the Synchron API that allows declarative specification of embedded applications. We suggest that this API is an improvement, in terms of expressivity, over the currently existing languages and libraries on embedded systems and provide illustrative examples to support this in Section 6. We also provide benchmarks on the Synchron runtime in Section 7. Next, we discuss the Synchron API in more detail.

## 3  The Synchron API

### 3.1  Synchronous Message-Passing and Events

We begin by looking at a standard synchronous message-passing API, like Hoare's CSP [17] -

```
1 spawn   : (() -> ()) -> ThreadId
2 channel : () -> Channel a
3 sendMsg : Channel a -> a -> ()
4 recvMsg : Channel a -> a
```

In the above type signatures, the type parameter, `a`, indicates a polymorphic type. The call to `spawn` allows the creation of a new process whose body is represented by the $(() \rightarrow ())$ type. The `channel ()` call creates a blocking *channel* along which a process can send or receive messages using `sendMsg` and `recvMsg` respectively. A channel blocks until a sender has a corresponding receiver and vice-versa. Multi-party communication in CSP is enabled using the *nondeterministic choice* operation that races between two sets of synchronous operations and chooses the one that succeeds first.

However, there is a fundamental conflict between procedural abstraction and the *choice* operation. Using a procedure to represent a complex protocol involving multiple *sends* and *receives* hides the critical operations over which a *choice* operation is run. On the other hand, exposing the individual *sends* and *receives* prevents the construction of protocols with strict message-ordering constraints (Appendix A). Reppy discusses this issue [27] in depth and provides a solution in Concurrent ML (CML), which is adopted by the Synchron API.

The central idea is to break the act of synchronous communication into two steps:

(i) Expressing the intent of communication as an *event*-value
(ii) Synchronising between the sender and receiver via the *event*-value

The first step above results in the creation of a type of value called an `Event`. An *event* is a first-class value in the language akin to the treatment of higher-order functions in functional languages. Reppy describes events as "first-class synchronous operations" [27]. Adapting this idea, the type signature of message sending and receiving in the Synchron API becomes :

> **send : Channel a → a → Event ()**
> **recv : Channel a → Event a**

Given a value of type `Event`, the second step of synchronising between processes and the consequent act of communication is accomplished via the `sync` operation, whose type is :

> **sync : Event a → a**

Intuitively, an equivalence can be drawn between the message passing in CSP and the CML-style message passing (as adopted in the Synchron API) using function composition:

```
sync . (send c)  ≡  sendMsg c
sync . recv      ≡  recvMsg
```

The advantage of representing communication as a first-class value is that event-based combinators can be used to build more elaborate communication protocols. In the same vein as higher-order functions like *function composition (.)* and *map*, events support composition via the following operators in the Synchron API :

> **choose : Event a → Event a → Event a**
> **wrap  : Event a → (a → b) → Event b**

The `choose` operator is equivalent to the *choice* operation in CSP and the `wrap` operator can be used to apply a post-synchronisation operation (of type $a \to b$). A large tree of *events* representing a communication protocol can be built in this framework as follows :

```
protocol : Event ()
protocol =
  choose (send c1 msg1)
         (wrap (recv c2) (λ msg2 -> sync (send c3 msg2)))
```

Using events, the above protocol involving multiple *sends* and *receives* was expressible as a procedural abstraction while still having the return type of `Event ()`. A consumer of the above protocol can further use the nondeterministic choice operator, `choose`, and *choose* among multiple protocols (see Appendix A). This combination of a composable functional programming style and CSP-inspired multiprocess program design allows this API to represent callback-based, state machine oriented programs in a declarative manner.

**Comparisons between Events and Futures.** The fundamental difference between events and futures is that of *deferred communication* and *deferred computation* respectively. A future aids in asynchronous computations by encapsulating a computation whose value is made available at a future time. On the other hand, an event represents deferred communication as a first-class entity in a language. Using the `wrap` combinator, it is possible to chain lambda functions capturing computations that should happen post-communication as well. However, events are fundamentally building blocks for communication protocols.

## 3.2 Input and Output

The I/O component of Synchron is deeply connected to the Synchron runtime. Hence, we mention parts of the low-level runtime implementation while describing the I/O interface.

In the Synchron API, I/O is expressed using the same events as are used for inter-process communication. Each I/O device is connected to the running program using a primitive we call `spawnExternal` as a hint that the programmer can think of, for example, an LED as a process that can receive messages along a channel. Each *external* process denotes an underlying I/O device that is limited to send and receive messages along one channel.

The `spawnExternal` primitive takes the channel to use for communication with software and a driver. It returns a "ThreadId" for symmetry with `spawn`.

> **spawnExternal : Channel a → Driver → ExternalThreadId**

The first parameter supplied to `spawnExternal` is a designated fixed channel along which the external process shall communicate. The second argument requires some form of identifier to uniquely identify the driver. This identifier for a driver tends to be architecture-dependent. For instance, when using low-level memory-mapped I/O, reads or writes to a memory address are used to communicate with a peripheral. So the unique memory address would be an identifier in that case. On the other hand, certain real-time operating system (such as `FreeRTOS` or `Zephyr`) can provide more high-level abstractions over a memory address. In the Synchron runtime, we number each peripheral in a monotonically increasing order, starting from 0. So our `spawnExternal` API becomes:

```
type DriverNo = Int
spawnExternal : Channel a -> DriverNo -> ExternalThreadId
```

In the rest of the paper, we will use suggestive names for drivers like `led0`, `uart1`, etc instead of plain integers for clarity. We have ongoing work to parse a file describing the target board/MCU system, automatically number the peripherals, and emit typed declarations, like $led0 = LED\ 0$, that can be used in the `spawnExternal` API.

To demonstrate the I/O API in the context of asynchronous drivers, we present a standard example of the *button-blinky* program. The program matches a button state to an LED so that when the button is down, the LED is on, otherwise the LED is off:

▪ **Listing 4** Button-Blinky using the Synchron API

```
butchan = channel ()
ledchan = channel ()

glowled i = sync (send ledchan i)

f : ()
f = let _ = sync (wrap (recv butchan) glowled) in f

main =
   let _ = spawnExternal butchan 0 in
   let _ = spawnExternal ledchan 1 in f
```

Listing 4 above spawns two hardware processes - an LED process and a button process. It then calls the function `f` which arrives at line 7 and waits for a button press. During the waiting period, the scheduler can put the process to sleep to save power. When the button interrupt arrives, the Synchron runtime converts the hardware interrupt to a software message and wakes up process `f`. It then calls the `glowled` function on line 4 that sends a switch-on message to the LED process and recursively calls `f` infinitely.

The above program represents an asynchronous, callback-based application in an entirely synchronous framework. The same application written in C, on top of the Zephyr OS, is more than 100 lines of callback-based code [11]. A notable aspect of the above program is the lack of any non-linear callback-handling mechanism.

The structure of this program resembles the event-loop pattern presented in Listing 3 but fixes all of its associated deficiencies - (1) all Synchron I/O operations are blocking; the runtime manages their optimal scheduling, not the programmer, (2) the internal plumbing related to interrupt-handling and queues are invisible to the programmer, (3) the program is highly extensible; adding a new interrupt handler is as simple as defining a new function.

### 3.3 Programming with Time

Real-time languages and frameworks generally provide a mechanism to override the fairness of a *fair* scheduler. A typical fair scheduler abstracts away the details of prioritising processes.

However, in a real-time scenario, a programmer wants to precisely control the response-time of certain operations. So the natural intuition for real-time C-extensions like FreeRTOS *Tasks* or languages like Ada is to delegate the scheduling control to the programmer by allowing them to attach a priority level to each process.

The priority levels involved decides the order in which a tie is broken by the scheduler. However, with a small fixed number of priority levels it is likely for several processes to end up with the same priority, leading the scheduler to order them fairly again within each level.

Another complication that crops up in the context of priorities is the *priority inversion problem* [33]. Priority inversion is a form of resource contention where a high-priority thread gets blocked on a resource held by a low-priority thread, thus allowing a medium priority thread to take advantage of the situation and get scheduled first. The outcome of this scenario is that the high-priority thread gets to run after the medium-priority thread, leading to possible program failures.

The Synchron API admits the *dynamic* prioritisation of processes, drawing inspiration from the TinyTimber kernel [23]. TinyTimber allows specifying a *timing window* expressed as a baseline and deadline time, and a scheduler can use this timing window to determine the runtime priority of a process. The timing window expresses the programmer's wish that the operation is started at the *earliest* on the baseline and *no later* than the deadline.

In Synchron, a programmer specifies a *timing window* (of the wall-clock time) during which they want message synchronisation, that is the rendezvous between message sender and receiver, to happen. We do this with the help of the timed-synchronisation operator, `syncT`, with type signature:

> **syncT : Time → Time → Event a → a**

Comparing the type signature of `syncT` with that of `sync` :

```
1 syncT : Time -> Time -> Event a -> a
2 sync  :                 Event a -> a
```

The two extra arguments to `syncT` specify a lower and upper bound on the *time of synchronisation* of an event. The two arguments to `syncT`, of type `Time`, express the relative times calculated from the current wall-clock time. The first argument represents the *relative baseline* - the earliest time instant from which the event synchronisation should begin. The second argument specifies the *relative deadline*, i.e. the latest time instant (starting from the baseline), by which the synchronisation should start. For instance,

```
1 syncT (msec 50) (msec 20) timed_ev
```

means that the event, *timed_ev*, should begin synchronisation at the earliest 50 milliseconds and the latest $50 + 20$ milliseconds from *now*. The *now* concept is based on a thread's local view of what time it is. This thread-local time ($T_{local}$) is always less than or equal to wall-clock time ($T_{absolute}$). When a thread is spawned, its thread-local time, $T_{local}$, is set to the wall-clock time, $T_{absolute}$.

While a thread is running, its local time is frozen and unchanged until the thread executes a timed synchronisation; a `syncT` operation where time progresses to $T_{local} + baseline$.

```
1  process1 _ =
2    let _ = s1 in -- T_local = 0
3    let _ = s2 in -- T_local = 0
4    let _ = syncT (msec 50) (usec 10) ev1 in
5    process1 () -- T_local = 50 msec
```

The above illustrates that the *untimed* operations `s1` and `s2` have no impact on a thread's view of what time it is. In essence, these operations are considered to take no time, which is a reference to *logical* time and not the physical time. Synchron shares this logical computational model with other systems such as the synchronous languages [7] and ChucK [37].

In practice, this assumption helps control jitter in the timing as long as the timing windows specified on the synchronisation is large enough to contain the execution time of `s1`, `s2`, the synchronisation step and the recursive call. Local and absolute time must meet up at key points for this approach to work. Without the two notions of time meeting, local time would drift away from absolute time in an unbounded fashion. For a practical implementation of `syncT`, a scheduler needs to meet the following requirements:

- The scheduler should provide a mechanism for overriding fair scheduling.
- The scheduler must have access to a wall-clock time source.
- A scheduler should attempt to schedule synchronisation such that local time meets up with absolute time at that instant.

We shall revisit these requirements in Section 5 when describing the scheduler within the Synchron runtime. Next, we shall look at a simple example use of `syncT`.

### Blinky

The well-known *blinky* example, shown in Listing 5, involves blinking an LED on and off at a certain frequency. Here we blink once every second.

**Listing 5** Blinky with `syncT`

```
1  not 1 = 0
2  not 0 = 1
3
4  ledchan = channel ()
5
6  sec  n = n * 1000000
7  usec n = n -- the unit-time in the Synchron runtime
8
9  foo : Int -> ()
10 foo val =
11   let _ = syncT (sec 1) (usec 1) (send ledchan val) in
12   foo (not val)
13
14 main = let _ = spawnExternal ledchan 1 in foo 1
```

In the above program, `foo` is the only software process, and there is one external hardware process for the LED driver that can be communicated with, using the `ledChan` channel. Line 11 is the critical part of the logic that sets the period of the program at 1 second, and the recursion at Line 12 keep the program alive forever. Appendix B shows the details of scheduling this program. We discuss a more involved example using `syncT` in Section 6.3.

## 4 Synchronisation Algorithms

The synchronous nature of message-passing is the foundation of the Synchron API. In this section, we describe the runtime algorithms, in an abstract form, that enable processes to synchronise. The Synchron runtime implements these algorithms, which drives the scheduling of the various software processes.

In Synchron, we synchronise on events. **Events**, in our API, fall into the categories of *base* events and *composite* events. The base events are `send` and `recv` and events created using `choose` are composite.

```
1  composite_event = choose (send c1 m1) (choose (send c2 m2) (send c3 m3))
```

From the API's point of view, composite events resemble a tree with base events in the leaves. However, for the algorithm descriptions here, we consider an event to be a *set* of base events. An implementation could impose an ordering on the base events that make up a composite event. Different orderings correspond to different event-prioritisation algorithms.

In the algorithm descriptions below, a **Channel** consists of two FIFO queues, one for `send` and one for `recv`. On these queues, process identities are stored. While blocked on a `recv` on a channel, that process' id is stored in the receive queue of that channel; likewise for `send` and the send-queue. Synchronous exchange of the message means that messages themselves do not need to be maintained on a queue.

Additionally, the algorithms below rely on there being a queue of processes that are ready to execute. This queue is called the `readyQ`. In the algorithm descriptions below, handling of `wrap` has been omitted. A function wrapped around an event specifies an operation that should be performed after synchronisation has been completed. Also, we abstract over the synchronisation of hardware events. As a convention, `self` used in the algorithms below refers to the process from which the `sync` operation is executed.

### 4.1 Synchronising events

The synchronisation algorithm, that performs the API operation `sync`, accepts a set of base events. The algorithm searches the set of events for a base event that has a sender or receiver blocked (ready to synchronise) and passes the message between sender and receiver. Algorithm 1 provides a high-level view of the synchronisation algorithm.

The first step in synchronisation is to see if there exists a synchronisable event in the set of base events. The *findSynchronisableEvent* algorithm is presented in Algorithm 2.

If the *findSynchronisableEvent* algorithm is unable to find an event that can be synchronised, the process initiating the synchronisation is blocked. The process identifier then gets added to all the channels involved in the base events of the set. This is shown in Algorithm 3.

After registering the process identifiers on the channels involved, the currently running process should yield its hold on the CPU, allowing another process to run. The next process to start running is found using the *dispatchNewProcess* algorithm in Algorithm 4.

When two processes are communicating, the first one to be scheduled will block as the other participant in the communication is not yet waiting on the channel. However, when *dispatchNewProcess* dispatches the second process, the *findSynchronisableEvent* function will return a synchronisable event and the *syncNow* operation does the actual message passing. The algorithm of *syncNow* is given in Algorithm 5 below.

**Algorithm 1** The synchronisation algorithm

**Data:** $event : Set$
$ev \leftarrow findSynchronisableEvent(event)$;
**if** $ev \neq \varnothing$ **then**
 | $syncNow(ev)$;
**else**
 | $block(event)$;
 | $dispatchNewProcess()$;
**end**

**Algorithm 2** The findSynchronisableEvent function

**Data:** $event : Set$
**Result:** A synchronisable event or $\varnothing$
**foreach** $e \in event$ **do**
 | **if** $e.baseEventType == SEND$ **then**
 |  | **if** $\neg isEmpty(e.channelNo.recvq)$ **then**
 |  |  | **return** $e$
 |  | **end**
 | **else if** $e.baseEventType == RECV$ **then**
 |  | **if** $\neg isEmpty(e.channelNo.sendq)$ **then**
 |  |  | **return** $e$
 |  | **end**
 | **else return** $\varnothing$;                                /* Impossible case */
**end**
**return** $\varnothing$ ;                            /* No synchronisable event found */

**Algorithm 3** The block function

**Data:** $event : Set$
**foreach** $e \in event$ **do**
 | **if** $e.baseEventType == SEND$ **then**
 |  | $e.channelNo.sendq.enqueue(self)$;
 | **else if** $e.baseEventType == RECV$ **then**
 |  | $e.channelNo.recvq.enqueue(self)$;
 | **else** Do nothing;                                /* Impossible case */
**end**

**Algorithm 4** The dispatchNewProcess function

**if** $readyQ \neq \varnothing$ **then**
 | $process \leftarrow dequeue(readyQ)$;
 | $currentProcess = process$;
**else**
 | *relinquish control to the underlying OS*
**end**

**Algorithm 5** The syncNow function

**Data:** A base-event value - *event*
**if** *event.baseEventType == SEND* **then**
  *receiver ← dequeue(event.channelNo.recvq)*;
  *deliverMSG(self, receiver, msg)* ;    /* pass msg from self to receiver */
  *readyQ.enqueue(self)*;
**else if** *event.baseEventType == RECV* **then**
  *sender ← dequeue(event.channelNo.sendq)*;
  *deliverMSG(sender, self, msg)* ;       /* pass msg from sender to self */
  *readyQ.enqueue(sender)*;
**else** Do nothing;                                          /* Impossible case */

## 4.2 Timed synchronisation of events

Now, we look at the *timed* synchronisation algorithms. Timed synchronisation is handled by a two-part algorithm - the first part (Algorithm 6) runs when a process is executing the `syncT` API operation, and the second part (Algorithm 7) is executed later, after the baseline time specified in the `syncT` call is reached.

These algorithms rely on there being an alarm facility based on absolute wall-clock time, which invokes Algorithm 7 at a specific time. The alarm facility provides the operation `setAlarm` used in the algorithms below. The algorithms also require a queue, `waitQ`, to hold processes waiting for their baseline time-point.

**Algorithm 6** The time function

**Data:** Relative Baseline = *baseline*, Relative Deadline = *deadline*
$T_{wakeup} = self.T_{local} + baseline$;
**if** *deadline == 0* **then**
  $T_{finish} = Integer.MAX$;       /* deadline = 0 implies no deadline */
**else**
  $T_{finish} = T_{wakeup} + deadline$;
**end**
$self.deadline = T_{finish}$;
$baseline_{absolute} = T_{absolute} + baseline$;
$deadline_{absolute} = T_{absolute} + baseline + deadline$;
$cond1 = T_{absolute} > deadline_{absolute}$;
$cond2 = (T_{absolute} \geq baseline_{absolute})\&\&(T_{absolute} \leq deadline_{absolute})$;
$cond3 = baseline < \epsilon$;              /* platform dependent small time period */
**if** $baseline == 0 \vee cond1 \vee cond2 \vee cond3$ **then**
  *readyQ.enqueue(currentThread)*;
  *dispatchNewProcess()*;
  *return*;
**end**
$setAlarm(T_{wakeup})$;
$waitQ.enqueue(self).orderBy(T_{wakeup})$;
*dispatchNewProcess()*;

The `handleAlarm` function in Algorithm 7 runs when an alarm goes off and, at that point,

makes a process from the waitQ ready for execution. When the alarm goes off, there could be a process running already that should either be preempted by a timed process with a tight deadline or be allowed to run to completion in case its deadline is the tightest. The other alternative is that there is no process running and the process acquired from the waitQ can immediately be made active.

**Algorithm 7** The handleAlarm function

**Data:** Wakeup Interrupt
$timedProcess \leftarrow dequeue(waitQ)$;
$T_{now} = timedProcess.baseline$;
$timedProcess.T_{local} = T_{now}$;
**if** $waitQ \neq \varnothing$ **then**
  $timedProcess_2 \leftarrow peek(waitQ)$;                /* Does not dequeue */
  $setAlarm(timedProcess_2.baseline)$;
**end**
**if** $currentProcess == \varnothing$;         /* No process currently running */
 **then**
  $currentProcess = timedProcess$;
**else**
  **if** $timedProcess.deadline < currentProcess.deadline$ **then**
    /* Preempt currently running process                    */
    $readyQ.enqueue(currentProcess)$;
    $currentProcess = timedProcess$;
  **else**
    /* Schedule timed process to run after currentProcess   */
    $readyQ.enqueue(timedProcess)$;
    $currentProcess.T_{local} = T_{now}$;      /* Avoids too much time drift */
  **end**
**end**

## 5 Implementation in SynchronVM

The algorithms of Section 4 are implemented within the Synchron runtime. The Synchron API and runtime are part of a larger programming platform that is the bytecode-interpreted virtual machine called SynchronVM, which builds on the work by Sarkar et al. [28].

The execution unit of SynchronVM is based on the Categorical Abstract Machine (CAM) [6]. CAM supports the cheap creation of closures, as a result of which SynchronVM can support a functional language quite naturally. CAM was chosen primarily for its simplicity and availability of pedagogical resources [16].

### 5.1 System Overview

Figure 4 shows the architecture of SynchronVM. The whole pipeline consists of three major components - (i) the frontend, (ii) the middleware and (iii) the backend.

**Frontend.** We support a statically-typed, eagerly-evaluated, Caml-like functional language on the frontend. The language comes equipped with Hindley-Milner type inference. The polymorphic types are monomorphised at compile-time. The frontend module additionally runs a lambda-lifting pass to reduce the heap-allocation of closures.

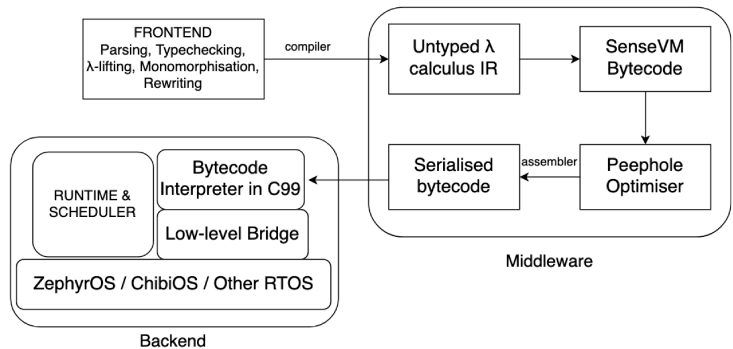

**Figure 4** The compiler and runtime for SynchronVM

**Middleware.** The frontend language gets compiled to an untyped lambda-calculus-based intermediate representation. This intermediate representation is then further compiled down to the actual bytecode that gets interpreted at run time. The generated bytecode emulates operations on a stack machine with a single environment register that holds the final value of a computation. This module generates specialised bytecodes that reduce the environment register-lookup using an operational notion called *r-free* variables described by Hinze [16]. On the generated bytecode, a peephole-optimisation pass applies further optimisations, such as $\beta$-reduction and *last-call optimisation* [16] (a generalisation of tail-call elimination).

**Backend.** The backend module is the principal component of the SynchronVM. It can be further classified into four components - (i) the bytecode interpreter, (ii) the runtime, (iii) a low-level *bridge* interface, and (iv) underlying OS/driver support.

*Interpreter.* The interpreter is written in C99 for portability. Currently, we use a total of 55 bytecodes operating on a stack machine.

*Runtime.* The runtime consists of a fixed-size stack with an environment register. A thread/process is emulated via the notion of a *context*, which holds a fixed-size stack, an environment register and a program counter to indicate the bytecode that is being interpreted. The runtime supports multiple but a *fixed* number of contexts, owing to memory constraints. A context holds two additional registers - one for the process-local clock ($T_{local}$) and the second to hold the deadline of that specific context (or thread).

The runtime has a garbage-collected heap to support closures and other composite values like tuples, algebraic data types, etc. The heap is structured as a list made of uniformly-sized tuples. For garbage collection, the runtime uses a standard *non-moving*, mark-and-sweep algorithm with the Hughes lazy-sweep optimisation [18].

*Low-level Bridge.* The runtime uses the low-level bridge interface functions to describe the various I/O-related interactions. It connects the runtime with the lowest level of the hardware. We discuss it in depth in Section 5.4.

*Underlying OS/drivers.* The lowest level of SynchronVM uses a real-time operating system that provides drivers for interacting with the various peripherals. The low-level is not restricted to use any particular OS, as shall be demonstrated in our examples using both the Zephyr-OS and ChibiOS.

### 5.1.1 Concurrency, I/O and Timing bytecode instructions

For accessing the operators of our programming interface as well as any general runtime-based operations, SynchronVM has a dedicated bytecode instruction - `CALLRTS n`, where `n` is a natural number to disambiguate between operations. Table 1 shows the bytecode operations corresponding to our programming interface.

| spawn   | CALLRTS 0 | recv   | CALLRTS 3 | spawnExternal | CALLRTS 6 |
|---------|-----------|--------|-----------|---------------|-----------|
| channel | CALLRTS 1 | sync   | CALLRTS 4 | wrap          | CALLRTS 7 |
| send    | CALLRTS 2 | choose | CALLRTS 5 | syncT         | CALLRTS 8; CALLRTS 4 |

**Table 1** Concurrency, I/O and Timing bytecodes

A notable aspect is the handling of the `syncT` operation, which gets compiled into a sequence of two instructions. The first instruction in the `syncT` sequence is `CALLRTS 8` which corresponds to Algorithm 6 in Section 4. When the process is woken up by Algorithm 7, the process program counter lands at the next instruction which is `CALLRTS 4` (`sync`).

## 5.2 Message-passing with events

All forms of communication and I/O in SynchronVM operate via synchronous message-passing. However, a distinct aspect of SynchronVM's message-passing is the separation between the *intent* of communication and the actual communication. A value of type `Event` indicates the intent to communicate.

An event-value, like a closure, is a concrete runtime value allocated on the heap. The fundamental event-creation primitives are `send` and `recv`, which Reppy calls base-event constructors [27]. The event composition operators like `choose` and `wrap` operate on these base-event values to construct larger events. When a programmer attempts to send or receive a message, an event-value captures the channel number on which the communication was desired. When this event-value is synchronised (Section 4), we use the channel number as an identifier to match between prospective senders and receivers. Listing 6 shows the heap representation of an event-value as the type `event_t` and the information that the event-value captures on SynchronVM.

**Listing 6** Representing an `Event` in SynchronVM

```
1  typedef enum {
2      SEND, RECV
3  } event_type_t;
4
5  typedef struct {
6    event_type_t e_type; //  8 bits
7    UUID channel_id;     //  8 bits
8  } base_evt_simple_t;
9
10 typedef struct {
11   base_evt_simple_t  evt_details; // stored with 16 bits free
12   cam_value_t wrap_func_ptr; // 32 bits
13 } base_event_t;
14
15
16 typedef struct {
17   base_event_t bev; // 32 bits
18   cam_value_t  msg; // 32 bits; NULL for recv
19 } cam_event_t;
20
21 typedef heap_index event_t;
```

An event is implemented as a linked list of base-events constructed by applications of the `choose` operation. Each element of the list captures (i) the message that is being sent or received, (ii) any function that is wrapped around the base-event using `wrap`, (iii) the channel being used for communication and (iv) an enum to distinguish whether the base-event is a `send` or `recv`. Fig 5 visualises an event upon allocation to the Synchron runtime's heap.

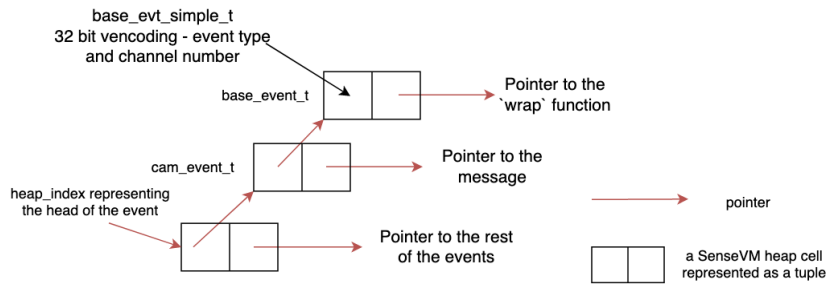

**Figure 5** An event on the SynchronVM heap

The linked-list, as shown above, is the canonical representation of an `Event`-type. It can represent any complex composite event. For instance, if we take an example composite event that is created using the base-events, $e_1, e_2, e_3$ and a wrapping function $wf_1$, it can always be rewritten to its canonical form.

```
choose e₁ (wrap (choose e₂ e₃) wf₁)

-- Rewrite to canonical form --

choose e₁ (choose (wrap e₂ wf₁) (wrap e₃ wf₁))
```

The `choose` operation can simply be represented as *cons*ing onto the event list.

## 5.3 The scheduler

SynchronVM's scheduler is a hyrbid of cooperative and preemptive scheduling. For applications that do not use `syncT`, the scheduler is cooperative in nature. Initially the threads are scheduled in the order that the main method calls them. For example,

```
main =
  let _ = spawn thread1 in
  let _ = spawn thread2 in
  let _ = spawn thread3 in ...
```

The scheduler orders the above in the order of `thread1` first, `thread2` next and `thread3` last. As the program proceeds, the scheduler relies on the threads to yield their control according to the algorithms of Section 4. When the scheduler is unable to find a matching thread for the currently running thread that is ready to synchronise the communication, it blocks the current thread and calls the *dispatchNewProcess*() function to run other threads (see Algorithm 1). On the other hand, when synchronisation succeeds, the scheduler puts the message-sending thread in the `readyQ` and the message-receiving thread starts running.

The preemptive behaviour of the scheduler occurs when using `syncT`. For instance, when a particular *untimed* thread is running and the baseline time of a timed thread has arrived, the scheduler then preempts the execution of the *untimed* thread and starts running the timed thread. A similar policy is also observed when the executing thread's deadline is later than a more urgent thread; the thread with the earliest deadline is chosen to be run at that instance. Algorithm 7 shows the preemptive components of the scheduler.

The SynchronVM scheduler also handles hardware driver interactions via message-passing. The structure that is used for messaging is shown below:

**Listing 7** A SynchronVM hardware message

```
typedef struct {
  uint32_t sender_id;
  uint32_t msg_type;
  uint32_t data;
  Time timestamp;
} svm_msg_t;
```

The `svm_msg_t` type contains a unique sender id for each driver that is the same as the number used in `spawnExternal` to identify that driver. The 32 bit `msg_type` field can be used to specify different meanings for the next field, the `data`. The `data` is a 32 bit word. The `timestamp` field of a message struct is a 64 bit entity, explained in detail in Section 5.5.

When the SynchronVM scheduler has all threads blocked, it uses a function pointer called `blockMsg`, which is passed to it by the OS that starts the scheduler, to wait for any interrupts from the underlying OS (more details in Section 5.4). Upon receiving an interrupt, the scheduler uses the SynchronVM runtime's `handleMsg` function to handle the corresponding message. The function internally takes the message and unblocks the thread for which the message was sent. The general structure of SynchronVM's scheduler is shown in Algorithm 8.

**Algorithm 8** The SynchronVM scheduler

**Data:** $blockMsg$ function pointer
$\forall threads$ set $T_{local} = T_{absolute}$;
$svm\_msg\_t\ msg$;
**while** $True$ **do**
  **if** $all\ threads\ blocked$ **then**
    $blockMsg(\&msg)$;           /* Relinquish control to OS */
    $handleMsg(msg)$;
  **else**
    $interpret(currentThread.PC)$;
  **end**
**end**

The $T_{local}$ clock is initialised for each thread when starting up the scheduler. Also notable is the `blockMsg` function that relinquishes control to the underlying OS, allowing it to save power. When the interrupt arrives, the `handleMsg` function unblocks certain thread(s) so that when the $if..then$ clause ends, in the following iteration the $else$ clause is executed and bytecode interpretation continues. We next discuss the low-level bridge connecting Synchron runtime to the underlying OS.

## 5.4 The Low-Level Bridge

The low-level bridge specifies two interfaces that should be implemented when writing peripheral drivers to use with SynchronVM. The first contains functions for reading and writing data synchronously to and from a driver. The second is geared towards interrupt-based drivers that asynchronously produce data.

The C-**struct** below contains the interface functions for reading and writing data to a driver as well as functions for checking the availability of data.

```c
typedef struct ll_driver_s{
  void *driver_info;
  bool is_synchronous;
  uint32_t (*ll_read_fun)(struct ll_driver_s *this, uint8_t*, uint32_t);
  uint32_t (*ll_write_fun)(struct ll_driver_s *this, uint8_t*, uint32_t);
  uint32_t (*ll_data_readable_fun)(struct ll_driver_s* this);
  uint32_t (*ll_data_writeable_fun)(struct ll_driver_s* this);

  UUID channel_id;
} ll_driver_t;
```

The `driver_info` field in the `ll_driver_t` struct can be used by a driver that implements the interface to keep a pointer to lower-level driver specific data. For interrupt-based drivers, this data will contain, among other things, an *OS interoperation* struct. These OS interoperation structs are shown further below. A boolean indicates whether the driver is synchronous or not. Next, the struct contains function pointers to the low-level driver's implementation of the interface. Lastly, a `channel_id` identifies the channel along which the driver is allowed to communicate with processes running on top of SynchronVM.

The `ll_driver_t` struct contains all the data associated with a driver's configuration in one place and defines a set of platform and driver independent functions for use in the runtime system, shown below:

```c
uint32_t ll_read(ll_driver_t *drv, uint8_t *data, uint32_t data_size);
uint32_t ll_write(ll_driver_t *drv, uint8_t *data, uint32_t data_size);
uint32_t ll_data_readable(ll_driver_t *drv);
uint32_t ll_data_writeable(ll_driver_t *drv);
```

The OS interoperation structs, mentioned above, are essential for drivers that asynchronously produce data. We show their Zephyr and ChibiOS versions below:

```c
typedef struct zephyr_interop_s {
  struct k_msgq *msgq;
  int (*send_message)(struct zephyr_interop_s* this, svm_msg_t msg);
} zephyr_interop_t;
```

```c
typedef struct chibios_interop_s {
  memory_pool_t *msg_pool;
  mailbox_t *mb;
  int (*send_message)(struct chibios_interop_s* this, svm_msg_t msg);
} chibios_interop_t;
```

In both cases, the struct contains the data that functions need to set up low-level message-passing between the driver and the OS thread running the SynchronVM runtime. Zephyr provides a message-queue abstraction that can take fixed-size messages, while ChibiOS supports a mailbox abstraction that receives messages that are the size of a pointer. Since ChibiOS mailboxes cannot receive data that is larger than a 32-bit word, a memory pool of messages is employed in that case. The structure used to send messages from the drivers is the already-introduced `svm_msg_t` struct, given in Listing 7.

The scheduler also needs to handle alarm interrupts from the wall-clock time subsystem, arising from the `syncT` operation. The next section discusses that component of SynchronVM.

### 5.5 The wall-clock time subsystem

Programs running on SynchronVM that make use of the timed operations rely on there being a monotonically increasing timer. The wall-clock time subsystem emulates this by implementing a 64bit timer that would take almost 7000 years to overflow at 84MHz frequency or about 36000 years at 16MHz. The timer frequency of 16MHz is used on the NRF52 board, while the timer runs at 84MHz on the STM32.

SynchronVM requires the implementation of the following functions for each of the platforms (such as ChibiOS and Zephyr) that it runs on :

```
1 bool     sys_time_init(void *os_interop);
2 Time     sys_time_get_current_ticks(void);
3 uint32_t sys_time_get_alarm_channels(void);
4 uint32_t sys_time_get_clock_freq(void);
5 bool     sys_time_set_wake_up(Time absolute);
6 Time     sys_time_get_wake_up_time(void);
7 bool     sys_time_is_alarm_set(void);
```

The timing subsystem uses the same OS interoperation structs as drivers do and thus has access to a communication channel to the SynchronVM scheduler. The interoperation is provided to the subsystem at initialisation using `sys_time_init`.

The key functionality implemented by the timing subsystem is the ability to set an alarm at an absolute 64-bit point in time. Setting an alarm is done using `sys_time_set_wake_up`. The runtime system can also query the timing subsystem to check if an alarm is set and at what specific time.

The low-level implementation of the timing subsystem is highly platform dependent at present. But on both Zephyr and ChibiOS, the implementation is currently based on a single 32-bit counter configured to issue interrupts at overflow, where an additional 32-bit value is incremented. Alarms can only be set on the lower 32-bit counter at absolute 32-bit values. Additional logic is needed to translate between the 64-bit alarms set by SynchronVM and the 32-bit timers of the target platforms. Each time the overflow interrupt happens, the interrupt service routine checks if there is an alarm in the next 32-bit window of time and in that case, enables a compare interrupt to handle that alarm. When the alarm interrupt happens, a message is sent to the SynchronVM scheduler in the same way as for interrupt based drivers, using the message queue or mailbox from the OS interoperation structure.

Revisiting the requirements for implementing `syncT` (Section 3.3), we find that our scheduler (1) provides a preemptive mechanism to override the fair scheduling, (2) has access to a wall-clock time source, and (3) implements an earliest-deadline-first scheduling policy that attempts to match the local time and the absolute time.

## 5.6 Porting SynchronVM to another RTOS

For porting SynchronVM to a new RTOS, one needs to implement - (1) the wall-clock time subsystem interface from Section 5.5, (2) the low-level bridge interface (Section 5.4) for each peripheral, and (3) a mailbox or message queue for communication between asynchronous drivers and the runtime system, required by the time subsystem.

Our initial platform of choice was ZephyrOS for its platform-independent abstractions. The first port of SynchronVM was on ChibiOS, where the wall-clock time subsystem was 254 lines of C-code. The drivers for LED, PWM, and DAC were about 100 lines of C-code each.

## 6 Case Studies

### Finite-State Machines with Synchron

We will begin with two examples of expressing state machines (involving callbacks) in the Synchron API. Our examples are run on the `NRF52840DK` microcontroller board containing four buttons and four LEDs. We particularly choose the button peripheral because its drivers have a callback-based API that typically leads to non-linear control-flows in programs.

## 6.1 Four-Button-Blinky

We build on the *button-blinky* program from Listing 4 presented in Section 3.2. The original program, upon a single button-press, would light up an LED corresponding to that press and switch off upon the button release. We now extend that program to produce a one-to-one mapping between four LEDs and four buttons such that button1 press lights up LED1, button2 lights up LED2, button3 lights up LED3 and button4 lights up LED4 (while the button releases switch off the corresponding LEDs).

The state machine of button-blinky is a standard two-state automaton that moves from the ON-state to OFF on button-press and vice versa. Now, for the four button-LED combinations, we have four state machines. We can combine them using the `choose` operator.

Listing 8 shows the important parts of the logic. The four state machines are declared in Lines 1 to 4, and their composition happens in Line 6 using the `choose` operator. See Appendix C.1 for the full program.

▌ **Listing 8** The Four-Button-Blinky program expressed in the Synchron API

```
press1 = wrap (recv butchan1) (λ x -> sync (send ledchan1 x))
press2 = wrap (recv butchan2) (λ x -> sync (send ledchan2 x))
press3 = wrap (recv butchan3) (λ x -> sync (send ledchan3 x))
press4 = wrap (recv butchan4) (λ x -> sync (send ledchan4 x))

anybutton = choose press1 (choose press2 (choose press3 press4))

program : ()
program = let _ = sync anybutton in program
```

## 6.2 A more intricate FSM

We now construct a more intricate finite-state machine involving intermediate states that can move to an error state if the desired state-transition buttons are not pressed. For this example a button driver needs to be configured to send only one message per button press-and-release. So there is no separate button-on and button-off signal but one signal per button.

In this FSM, we glow the LED1 upon consecutive presses of button1 and button2. We use the same path to turn LED1 off. However, if a press on button1 is followed by a press of button 1 or 3 or 4, then we move to an error state indicated by LED3. We use the same path to switch off LED3. In a similar vein, consecutive presses of button3 and button4 turns on LED2 and button3 followed by button 1 or 2 or 3 turns on the error LED - LED3. Fig. 6 shows the FSM diagram of this application, omitting self-loops in the OFF state.

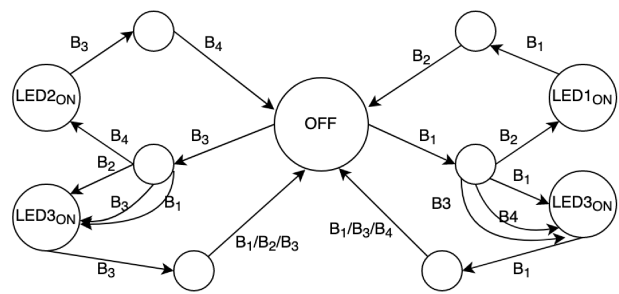

▌ **Figure 6** A complex state machine

Listing 9 shows the central logic expressing the FSM of Fig 6 in the Synchron API (see Appendix C.2 for the full program). This FSM can be viewed as a composition of two separate finite state machines, one on the left side of the OFF state involving LED2 and LED3 and one on the right side involving LED1 and LED3. Once again, we use the `choose` operator to compose these two state machines.

**Listing 9** The complex state machine running on the SynchronVM

```
errorLed x = ledchan3

fail1ev = choose (wrap (recv butchan1) errorLed)
         (choose (wrap (recv butchan3) errorLed)
                 (wrap (recv butchan4) errorLed))

fail2ev = choose (wrap (recv butchan1) errorLed)
         (choose (wrap (recv butchan2) errorLed)
                 (wrap (recv butchan3) errorLed))

led1Handler x =
   sync (choose (wrap (recv butchan2) (\x -> ledchan1)) fail1ev)

led2Handler x =
   sync (choose (wrap (recv butchan4) (\x -> ledchan2)) fail2ev)

led : Int -> ()
led state =
  let fsm1 = wrap (recv butchan1) led1Handler in
  let fsm2 = wrap (recv butchan3) led2Handler in
  let ch = sync (choose fsm1 fsm2) in
  let _ = sync (send ch (not state)) in
  led (not state)
```

In Listing 9, the `led1Handler1` and `ledHandler2` functions capture the intermediate states after one button press, when the program awaits the next button press. The error states are composed using the `choose` operator in the functions `fail1ev` and `fail2ev`.

The compositional nature of our framework is visible in line no. 21 where we compose the two state machines, `fsm1` and `fsm2`, using the `choose` operator. Synchronising on this composite event returns the LED channel (demonstrating a higher-order approach) on which the process should attempt to write. This program is notably a highly callback-based, reactive program that we have managed to represent in an entirely synchronous framework.

### 6.3   A soft-realtime music playing example

We present a soft-realtime music playing exercise from a Real-Time Systems course, expressed using the Synchron API. We choose the popular nursery rhyme - "Twinkle, Twinkle, Little Star". The program plays the tune repeatedly until it is stopped.

The core logic of the program involves periodically writing a sequence of 1's and 0's to a DAC driver. However, to make the produced note sound musical to the human ear, the *periodic rate* at which our process writes to the DAC driver is important, and this is where the real-time aspect of the application comes in. The human ear recognises a note produced at a certain frequency as a musical note. Our sound is generated at the 196Hz G3 music key.

Listing 10 shows the principal logic of the program expressed using the Synchron API. Note that we use `syncT` to describe a new temporal combinator `after` that determines the periodicity of this program. The list `twinkle` (line 17) holds the 28 notes in the twinkle song and the list `durations` (line 18) provides the length of each note. Appendix C provides full details of the program.

**Listing 10** The *Twinkle, Twinkle* tune expressed using the Synchron API

```
msec t = t * 1000
usec t = t
```

```
 3 after t ev = syncT t 0 ev
 4
 5 -- note frequencies
 6 g = usec 2551
 7 a = usec 2273
 8 b = usec 2025
 9 c = usec 1911
10 d = usec 1703
11 e = usec 1517
12
13 hn = msec 1000 -- half note
14 qn = msec  500 -- quarter note
15
16 twinkle, durations : List Int
17 twinkle   = [ g,  g,  d,  d,  e,  e,  d.... ] -- 28 notes
18 durations = [qn, qn, qn, qn, qn, qn, hn.... ]
19
20 dacC  = channel ()
21 noteC = channel ()
22
23 playerP : List Int -> List Int -> Int -> () -> ()
24 playerP melody nt n void =
25    if (n == 29)
26    then let _ = after (head nt) (send noteC (head twinkle)) in
27         playerP (tail twinkle) durations 2 void
28    else let _ = after (head nt) (send noteC (head melody)) in
29         playerP (tail melody) (tail nt) (n + 1) void
30
31 tuneP : Int -> Int -> () -> ()
32 tuneP timePeriod vol void =
33   let newtp =
34       after timePeriod (choose (recv noteC)
35                                (wrap (send dacC (vol * 4095))
36                                      (λ _ -> timePeriod))) in
37   tuneP newtp (not vol) void
38
39 main =
40    let _ = spawnExternal dacC 0 in
41    let _ = spawn (tuneP (head twinkle) 1) in
42    let _ = spawn (playerP (tail twinkle) durations 2) in ()
```

The application consists of two software processes and one external hardware process. We use two channels - `dacC` to communicate with the DAC and `noteC` to communicate between the two software processes. Looking at what each software process is doing -

*playerP*. This process runs at the rate of a note's length. For a quarter note it wakes up after 500 milliseconds (1000 msecs for a half note), traverses the next element of the `twinkle` list and sends it along the `noteC` channel. It circles back after completing all 28 notes.

*tuneP*. This process creates the actual sound. Its running rate varies depending on the note that is being played. For instance, when playing note C, it will write to the DAC at a rate of 1911 microseconds-per-write. However, upon receiving a new value along `noteC`, it changes its write frequency to the new value resulting in changing the note of the song.

## 7 Benchmarks

### 7.1 Interpretive overhead measurements

We characterise the overhead of executing programs on top of SynchronVM, compared to running them directly on either Zephyr or ChibiOS, by implementing *button-blinky* directly on top of these operating systems and measuring the response-time differences.

The button-blinky program copies the state of a button onto an LED, something that could be done very rapidly at a large CPU utilization cost by continuously polling the button state and writing it to the LED. Instead, the Zephyr and ChibiOS implementations are interrupt-based and upon receiving a button interrupt (for either button press or release), send a message to a thread that is kept blocking until such messages arrive. When the thread receives a message indicating a button down or up, it sets the LED to on or off. This approach keeps the low-level implementation in Zephyr and ChibiOS similar to SynchronVM and indicates the interpretive and other overheads in SynchronVM.

The data in charts presented here is collected using an STM32F4 microcontroller based testing system connected to either the NRF52 or the STM32F4 system under test (SUT). The testing system provides the stimuli, setting the GPIO (button) to either active or inactive and measures the time it takes for the SUT to respond on another GPIO pin (symbolising the LED). The testing system connects to a computer displaying a GUI and generates the plots used in this paper. Each plot places measured response times into buckets of similar time, and shows the number of samples falling in each bucket as a vertical bar. Each bucket is labelled with the average time of the samples it contains.

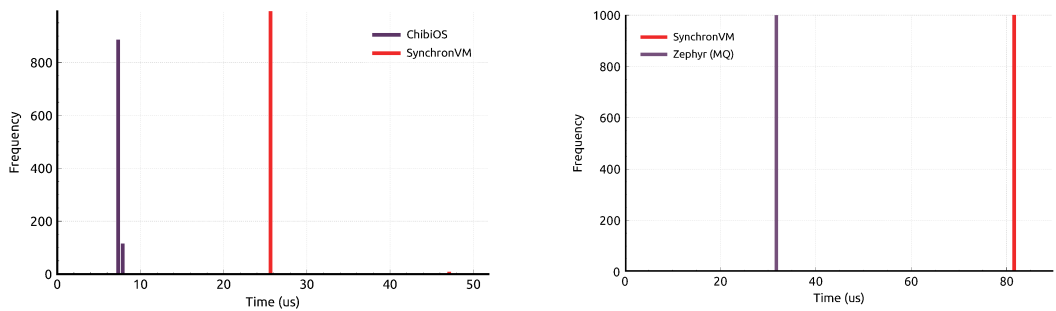

**(a)** Response time comparison between a C-code implementation using ChibiOS against the same program on SynchronVM (running on ChibiOS). Data obtained on the STM32F4 microcontroller. Uses 1000 samples.

**(b)** Response time comparison between a C-code implementation using Zephyr OS against the same program on SynchronVM (running on Zephyr). Data obtained on the NRF52 microcontroller. Uses 1000 samples.

**Figure 7** Button-blinky response times comparison between C and SynchronVM

Fig. 7a shows the SynchronVM response time in comparison to the implementation of the program running on ChibiOS using its mailbox abstraction (MB). There the overhead is about 3x. Fig. 7b compares response times for SynchronVM and the Zephyr message queue based implementation (MQ), and shows an overhead of 2.6x.

## 7.2 Effects of Garbage Collection

This experiment measures the effects of garbage collection on response time by repeatedly running 10000 samples test for different heap-size configurations of SynchronVM. A smaller heap should lead to more frequent interactions with the garbage collector, and the effects of the garbage collector on the response time should magnify.

As a smaller heap is used, the number of outliers should increase if the outliers are due to garbage collection. The following table shows the number of outliers at each size configuration for the heap used, and there is an indication that GC is the cause of outliers.

| Heap size (bytes) | 256 | 512 | 1024 | 2048 | 4096 | 8192 |
|---|---|---|---|---|---|---|
| Outliers NRF52 on Zephyr | 3334 | 1429 | 811 | 491 | 0 | 81 |
| Outliers STM32 on ChibiOs | 3339 | 1430 | 810 | 491 | 0 | 80 |

Figures 8 and 9 show the response-time numbers across the heap sizes of 8192, 4096, 2048, 1024, 512 and 256 bytes. A general observable trend is that as the heap size decreases and GC increases, the response time numbers hover towards the farther end of the X-axis. This trend is most visible for the heap size of 256 bytes, which is our smallest heap size. Note that we cannot collect enough sample data for response-time if we switch off the garbage collector (as a reference value), as the program would very quickly run out of memory and terminate.

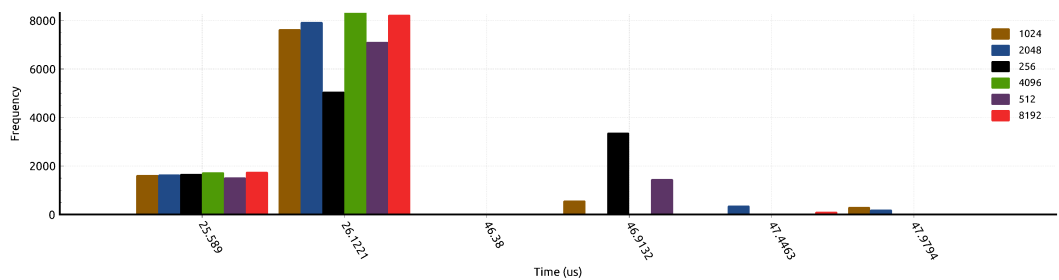

**Figure 8** Response time measurements at different sizes of the heap to identify effects of garbage collection. This data is collected on the STM32F4 microcontroller running SynchronVM on top of ChibiOS. Each bucket size is approx 0.533us. Uses 10000 samples.

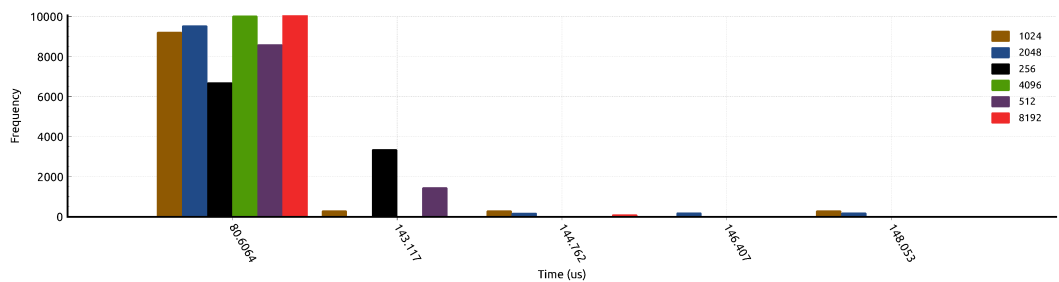

**Figure 9** Response time measurements at different sizes of the heap to identify effects of garbage collection. This data is collected on the NRF52 microcontroller running SynchronVM on top of the Zephyr OS. Each bucket size is approx 1.65us. Uses 10000 samples.

## 7.3 Memory Footprint

SynchronVM resides on the Flash memory of a microcontroller. On Zephyr, a tiny C application occupies 17100 bytes, whereas the same SynchronVM application occupies 49356 bytes, which gives the VM's footprint as 32256 bytes. For ChibiOS, the C application takes 18548 bytes, while the SynchronVM application takes 53868 bytes. Thus, SynchronVM takes 35320 bytes in this case. Hence, we can estimate SynchronVM's rough memory footprint at 32 KB, which will grow with more drivers.

## 7.4 Power Usage

Fig. 10 shows the power usage of the NRF52 microcontroller running the button-blinky program for three implementations. The first is a polling version of the program in C. The second program uses a callback-based version of button-blinky [11]. The last program is Listing 4 running on SynchronVM. The measurements are made using the Ruideng UM25C ammeter. We collect momentary readings from the ammeter after the value has stabilised. Notable in Fig. 10 is the polling-based C implementation's use of 0.0175 Watts of power in a button-off state, whereas SynchronVM consumes five times less power (0.0035 Watts).

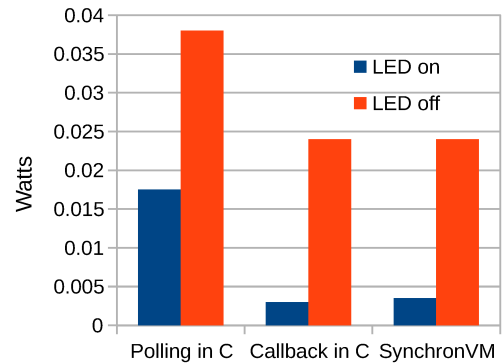

**Figure 10** Power usage measured on the NRF52 microcontroller

This is comparable to the callback-based C implementation's use of 0.003 Watts. Integrating the power usage over time will likely make the difference between SynchronVM and the callback-based C version more noticeable. However, we believe that the simplicity and declarative nature of the Synchron-based code provide a fair tradeoff.

## 7.5 Jitter and Precision

Jitter can be defined as the deviation from true periodicity of a presumably periodic signal, often in relation to a reference clock signal. We want to evaluate how our claims of `syncT` reducing jitter pans out in practice.

Listing 11 below is written in a naive way to illustrate how jitter manifests in programs. Figure 11a shows what the oscilloscope draws, set to persistent mode drawing while sampling the signal from the Raspberry Pi outputs.

The Raspberry Pi program reads the status of a GPIO pin and then inverts its state back to that same pin. The program then goes to sleep using `usleep` for 400us. The goal frequency was 1kHz and sleeping for 400us here gave a roughly 1.05kHz signal. The more expected sleep time of 500us to generate a 1kHz signal led, instead, to a much lower frequency. So, the 400us value was found experimentally.

```
1  while (1) {
2    uint32_t state = GPIO_READ(23);
3    if (state) {
4      GPIO_CLR(23);
5    } else {
6      GPIO_SET(23);
7    }
8    usleep(400);
9  }
10 // main method and other setup
       elided
```

**Listing 11** Raspberry Pi C code

```
11 ledchan = channel ()
12
13 foo : Int -> ()
14 foo val =
15   let _ = syncT 500 0 (send
       ledchan val)
16   in foo (not val)
17
18 main =
19   let _ = spawnExternal ledchan 1
20   in foo 1
```

**Listing 12** SynchronVM 1KHz wave code

Listing 12 shows the same 1kHz frequency generator for SynchronVM. Note that, in this case, specifying a baseline of 500us led to a 1kHz wave (Fig. 11b). In comparison, a 400us period in Listing 11 generated a roughly 1kHz wave, owing to additional delays of the system.

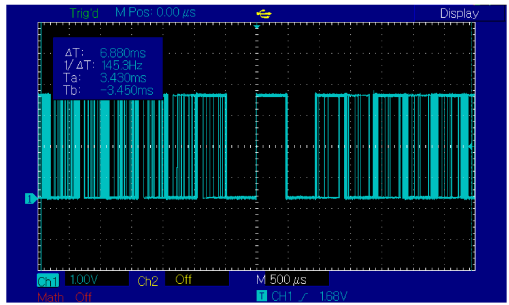

(a) Illustrating the amount of jitter on the square wave generated from the Raspberry Pi by setting the oscilloscope display in persistent mode.

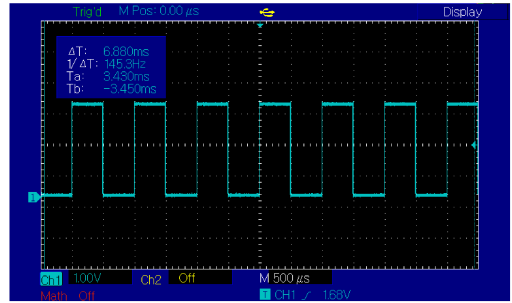

(b) A 1kHz square wave generated using SynchronVM running on the STM32F4 with no jitter

**Figure 11** A 1 kHz frequency generator on the Raspberry PI (in C) and STM32 (Synchron)

## 7.6 Load Test

The SynchronVM program in the previous section could produce a 1kHz-wave with no jitter. However, the only operation that the program did was produce the square wave. In this section, we want to test how much computational load can be performed by Synchron while producing the square wave. We emulate the workload using the following program.

```
21 load i n =
22   let _ = fib_tailrec n in
23   let _ = syncT 8000 0 (send
        ledchan i)
24   in load (not i)
```

```
25 loop i a b n =
26   if i == n then a
27   else loop (i+1) (b) (a+b) n
28
29 fib_tailrec n = loop 0 0 1 n
```

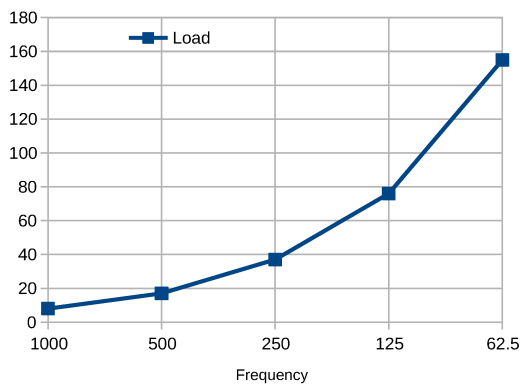

**Figure 12** Load testing SynchronVM with the nth fibonacci number function

At a given frequency, it is possible to calculate only up to a certain Fibonacci number while generating the square wave at the desired frequency. For example, when generating a 62.5 Hz wave, it is only possible to calculate up to the 155th Fibonacci number. If the 156th number is calculated, the wave frequency drops below 62.5 Hz.

Fig. 12 plots the nth Fibonacci numbers that can be calculated against the square wave frequencies that get generated without jitters. Our implementation of `fib_tailrec` involves 2 addition operations, 1 equality comparison and 1 recursive call. So, calculating the 155th Fibonacci number involves $155 * 4 = 620$ major operations. The trend shows that the load capacity of SynchronVM grows linearly as the desired frequency of the square wave is halved.

## 7.7 Music Program Benchmarks

We now provide some benchmarks on the music program from Section 6.3. Figure 13 shows CPU usage, average time it takes to allocate data and total time spent doing allocations in a 1 minute window. The values used in the chart come from the second minute of running the music application. The values from the first minute of execution are discarded as those would include the startup phase of the system. The amount of heap made available to the runtime system is varied from a roomy 16384 bytes down to 3072 bytes.

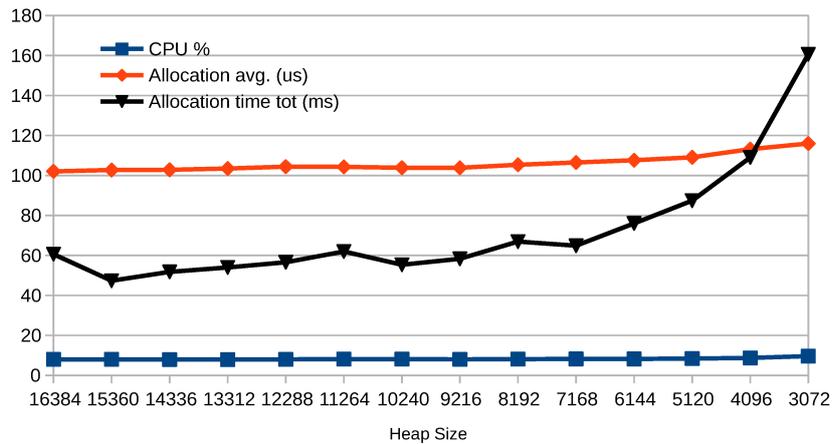

**Figure 13** CPU usage and allocation trends over a 1 minute window for Listing 10

The sweep phase of our garbage collector is intertwined with the allocations phase. Hence, instead of showing the GC time, the chart shows statistics related to all allocations that take a measurable amount of time using the ChibiOS 10KHz system timer. All allocations taking less than 100us are left out of the statistics (and not counted towards averaging).

The data in Fig. 13 shows that CPU usage of the music application is pretty stable at around 8 percent over the one minute window. It increases slightly for the very small heap sizes and ends up at nearly 10 percent at the smallest heap size that can house the program.

In terms of allocation, the average time of an allocation (in usecs) increases when the probability of a more expensive allocation increases, which in turn increases with small heap sizes. In the last data series, the total amount of time spent in allocations (in msecs) grows considerably as the heap size drops below 7168 bytes - an indicator of increased GC activity.

**Programming Complexity**. This music application ports a Real-Time Systems course exercise written in C using the TinyTimber kernel [21]. The TinyTimber-based C program (excluding the kernel) is around 600 lines of code. In comparison, our entire program (Appendix D), along with the library functions, stands at 74 lines. The most time-consuming part of the port was modifying the core logic in terms of message-passing. A major gain is that the interrupt-handling and other I/O management routines are invisible to the programmer. The user-defined timing operator, `after`, further enables the concision of the program.

## 7.8   Discussion

Our benchmarks, so far, show promising results for power, memory, and CPU usage. However, SynchronVM's response time is 2-3x times slower than native C code, which needs improvement. We attribute the slowness to the CAM-based execution engine, which we hope to mitigate by moving to a ZAM-based machine [20].

Our synthetic load test (Fig. 12) indicates that the VM can support around 150 operations for applications that operate around 250Hz (such as humanoid balance bots [9], autonomous vehicle platforms [36]). Our music program falls in the range of 200-500 Hz, and SynchronVM could sustain that frequency without introducing any jitter. There exist other *untimed*, aperiodic applications with much lower frequencies where SynchronVM could be applicable. Examples include smart home applications [34], monitoring systems [14], etc.

The Synchron API chooses a synchronous message-passing model, in contrast, with actor-based systems like Erlang that support an asynchronous message-passing model with each process containing a mailbox. We believe that a synchronous message-passing policy is better suited for embedded systems for the following reasons:

1. Embedded systems are highly memory-constrained, and asynchronous send semantics assume the *unboundedness* of an actor's mailbox, which is a poor assumption in the presence of memory constraints. Once the mailbox becomes full, message-sending becomes blocking, which is already the default semantics of synchronous message-passing.
2. Acknowledgement is implicit in synchronous message-passing systems, in contrast to explicit message acknowledgement in asynchronous systems that leads to code bloat. Additionally, if a programmer forgets to remove acknowledgement messages from an actor's mailbox, it leads to memory leaks.

## 8 Limitations and Future Work

In this section, we propose future work to improve the Synchron API and runtime.

### 8.1 Synchron API limitation

**Deadline miss API.** Currently, the Synchron API cannot represent actions that should happen if a task were to miss its deadline. We envision adapting the negative acknowledgement API of CML to represent missed-deadline handlers for Synchron.

### 8.2 SynchronVM limitations

**Memory management**. A primary area of improvement is upgrading our stop-the-world mark and sweep garbage collector and investigating real-time garbage collectors like Schism [25]. Another relevant future work would be investigating static memory-management schemes like regions [32] and techniques combining regions with GC [15].

**Interpretation overhead**. A possible approach to reducing our interpretation overhead could be pre-compiling our bytecode to machine code (AOT compilation). Similarly, dynamic optimization approaches like JITing could be an area of investigation.

**Priority inversions**. Although TinyTimber-style dynamic priorities might reduce priority inversion occurrences, they can still occur on the SynchronVM. Advanced approaches like priority inheritance protocols [29] need to be experimented with on our scheduler.

### 8.3 General runtime limitations

- Power efficiency and lifetime while operating from a small battery is challenging for a byte-code interpreting virtual machine.
- Safety-critical, hard real-time systems remain out of reach with a bytecode-interpreted and garbage collected virtual machine.

## 9 Related Work

Among functional languages running on microcontrollers, there exists OCaml running on OMicroB [35], Scheme running on Picobit [31] and Erlang running on AtomVM [4]. Synchron differs from these projects in the aspect that we identify certain fundamental characteristics of embedded systems and accordingly design an API and runtime to address those demands. As a result, our programming interface aligns more naturally to the requirements of an embedded systems application, in contrast with general-purpose languages like Scheme.

The Medusa [2] language and runtime is the inspiration behind our uniform framework of concurrency and I/O. Medusa, however, does not provide any timing based APIs, and their message-passing framework is based on the actor model (See Section 7.8).

In the real-time space, a safety-critical VM that can provide hard real-time guarantees on Real-Time Java programs is the FijiVM [26] implementation. A critical innovation of the project was the Schism real-time garbage collector [25], from which we hope to draw inspiration for future work on memory management.

RTMLton [30] is another example of a real-time project supporting a general-purpose language like SML. RTMLton adapts the MLton runtime [38] with ideas from FijiVM to enable handling real-time constraints in SML. CML is available as an SML library, so RTMLton provides access to the event framework of CML but lacks the uniform concurrency-I/O model and the `syncT` operator of Synchron.

The Timber language [5] is an object-oriented language that inspired the `syncT` API of Synchron. Timber was designed for hard real-time scenarios; related work on estimating heap space bounds [19] could perhaps benefit our future research.

The WebAssembly project (WASM) has spawned sub-projects like WebAssembly Micro Runtime (WAMR) [1] that allows running languages that compile to WASM on microcontrollers. Notable here is that while several general-purpose languages, like JavaScript, can execute on ARM architectures by compiling to WebAssembly, they lack the native support for the concurrent, I/O-bound, and timing-aware programs that is naturally provided by our API and its implementation. Reactive extensions of Javascript, like HipHop.js [3], are being envisioned to be used for embedded systems.

Another related line of work is embedding domain-specific languages like Ivory [10] and Copilot [24] in Haskell to generate C programs that can run on embedded devices. This approach differs from ours in the aspect that two separate languages dictate the programming model of an EDSL - the first being the DSL itself and the second being the host language (Haskell). We assess that having a single language (like in Synchron) provides a more uniform programming model to the programmer. However, code-generating EDSLs have very little runtime overheads and, when fully optimised, can produce high performance C.

## 10  Conclusion

In this paper, we have presented Synchron - an API and runtime for embedded systems. The API is implemented as a virtual machine called SynchronVM. We identified three essential characteristics of embedded applications, namely being concurrent, I/O–bound, and timing-aware. Correspondingly, our API is designed to address all three concerns. Our evaluations, conducted on the STM32 and NRF52 microcontrollers, show encouraging results for power, memory and CPU usage of the SynchronVM. Our response time numbers are within the range of 2-3x times that of native C programs, which we envision being improved by moving to a register-based execution engine and by using smarter memory-management strategies. We have additionally demonstrated the expressivity of our API through state machine-based examples, commonly found in embedded systems. Finally, we illustrated our timing API by expressing a soft real-time application, and we expect future theoretical investigations on the worst-case execution time and schedulability analysis on SynchronVM.

—— **References** ——

1   WAMR - WebAssembly Micro Runtime, 2019. URL: https://github.com/bytecodealliance/wasm-micro-runtime.
2   Thomas W. Barr and Scott Rixner. Medusa: Managing Concurrency and Communication in Embedded Systems. In Garth Gibson and Nickolai Zeldovich, editors, *2014 USENIX Annual Technical Conference, USENIX ATC '14, Philadelphia, PA, USA, June 19-20, 2014*, pages 439–450. USENIX Association, 2014. URL: https://www.usenix.org/conference/atc14/technical-sessions/presentation/barr.
3   Gérard Berry and Manuel Serrano. Hiphop.js: (A)Synchronous reactive web programming. In Alastair F. Donaldson and Emina Torlak, editors, *Proceedings of the 41st ACM SIGPLAN International Conference on Programming Language Design and Implementation, PLDI 2020, London, UK, June 15-20, 2020*, pages 533–545. ACM, 2020. doi:10.1145/3385412.3385984.
4   Davide Bettio. AtomVM, 2017. URL: https://github.com/bettio/AtomVM.
5   Andrew P Black, Magnus Carlsson, Mark P Jones, Richard Kieburtz, and Johan Nordlander. Timber: A programming language for real-time embedded systems. Technical report, OGI School of Science and Engineering, Oregon Health and Sciences University, Technical Report CSE 02-002. April 2002, 2002.
6   Guy Cousineau, Pierre-Louis Curien, and Michel Mauny. The Categorical Abstract Machine. In Jean-Pierre Jouannaud, editor, *Functional Programming Languages and Computer Architecture, FPCA 1985, Nancy, France, September 16-19, 1985, Proceedings*, volume 201 of *Lecture Notes in Computer Science*, pages 50–64. Springer, 1985. doi:10.1007/3-540-15975-4\_29.
7   Robert de Simone, Jean-Pierre Talpin, and Dumitru Potop-Butucaru. The Synchronous Hypothesis and Synchronous Languages. In Richard Zurawski, editor, *Embedded Systems Handbook*. CRC Press, 2005. doi:10.1201/9781420038163.ch8.
8   Adam Dunkels, Oliver Schmidt, Thiemo Voigt, and Muneeb Ali. Protothreads: simplifying event-driven programming of memory-constrained embedded systems. In Andrew T. Campbell, Philippe Bonnet, and John S. Heidemann, editors, *Proceedings of the 4th International Conference on Embedded Networked Sensor Systems, SenSys 2006, Boulder, Colorado, USA, October 31 - November 3, 2006*, pages 29–42. ACM, 2006. doi:10.1145/1182807.1182811.
9   Ahmed Elhasairi and Alexandre N. Pechev. Humanoid Robot Balance Control Using the Spherical Inverted Pendulum Mode. *Frontiers Robotics AI*, 2:21, 2015. doi:10.3389/frobt.2015.00021.
10  Trevor Elliott, Lee Pike, Simon Winwood, Patrick C. Hickey, James Bielman, Jamey Sharp, Eric L. Seidel, and John Launchbury. Guilt free ivory. In Ben Lippmeier, editor, *Proceedings of the 8th ACM SIGPLAN Symposium on Haskell, Haskell 2015, Vancouver, BC, Canada, September 3-4, 2015*, pages 189–200. ACM, 2015. doi:10.1145/2804302.2804318.
11  Zephyr examples. Zephyr button blinky, 2021. URL: https://pastecode.io/s/szpf673u.
12  The Linux Foundation. Zephyr RTOS. https://www.zephyrproject.org/. Accessed 2021-11-28.
13  Damien George. Micropython, 2014. URL: https://micropython.org/.
14  R. Kingsy Grace and S. Manju. A Comprehensive Review of Wireless Sensor Networks Based Air Pollution Monitoring Systems. *Wirel. Pers. Commun.*, 108(4):2499–2515, 2019. doi:10.1007/s11277-019-06535-3.
15  Niels Hallenberg, Martin Elsman, and Mads Tofte. Combining Region Inference and Garbage Collection. In Jens Knoop and Laurie J. Hendren, editors, *Proceedings of the 2002 ACM SIGPLAN Conference on Programming Language Design and Implementation (PLDI), Berlin, Germany, June 17-19, 2002*, pages 141–152. ACM, 2002. doi:10.1145/512529.512547.
16  Ralf Hinze. The Categorical Abstract Machine: Basics and Enhancements. Technical report, University of Bonn, 1993.
17  C. A. R. Hoare. Communicating Sequential Processes. *Commun. ACM*, 21(8):666–677, 1978. doi:10.1145/359576.359585.

**18**   R John M Hughes. A semi-incremental garbage collection algorithm. *Software: Practice and Experience*, 12(11):1081–1082, 1982.

**19**   Martin Kero, Pawel Pietrzak, and Johan Nordlander. Live Heap Space Bounds for Real-Time Systems. In Kazunori Ueda, editor, *Programming Languages and Systems - 8th Asian Symposium, APLAS 2010, Shanghai, China, November 28 - December 1, 2010. Proceedings*, volume 6461 of *Lecture Notes in Computer Science*, pages 287–303. Springer, 2010. `doi:10.1007/978-3-642-17164-2\_20`.

**20**   Xavier Leroy. *The ZINC experiment: an economical implementation of the ML language.* PhD thesis, INRIA, 1990.

**21**   Per Lindgren, Johan Eriksson, Simon Aittamaa, and Johan Nordlander. TinyTimber, Reactive Objects in C for Real-Time Embedded Systems. In *2008 Design, Automation and Test in Europe*, pages 1382–1385, 2008. `doi:10.1109/DATE.2008.4484933`.

**22**   Tommi Mikkonen and Antero Taivalsaari. Web Applications - Spaghetti Code for the 21st Century. In Walter Dosch, Roger Y. Lee, Petr Tuma, and Thierry Coupaye, editors, *Proceedings of the 6th ACIS International Conference on Software Engineering Research, Management and Applications, SERA 2008, 20-22 August 2008, Prague, Czech Republic*, pages 319–328. IEEE Computer Society, 2008. `doi:10.1109/SERA.2008.16`.

**23**   Johan Nordlander. *Programming with the TinyTimber kernel*. Luleå tekniska universitet, 2007.

**24**   Lee Pike, Alwyn Goodloe, Robin Morisset, and Sebastian Niller. Copilot: A Hard Real-Time Runtime Monitor. In Howard Barringer, Yliès Falcone, Bernd Finkbeiner, Klaus Havelund, Insup Lee, Gordon J. Pace, Grigore Rosu, Oleg Sokolsky, and Nikolai Tillmann, editors, *Runtime Verification - First International Conference, RV 2010, St. Julians, Malta, November 1-4, 2010. Proceedings*, volume 6418 of *Lecture Notes in Computer Science*, pages 345–359. Springer, 2010. `doi:10.1007/978-3-642-16612-9\_26`.

**25**   Filip Pizlo, Lukasz Ziarek, Petr Maj, Antony L. Hosking, Ethan Blanton, and Jan Vitek. Schism: fragmentation-tolerant real-time garbage collection. In Benjamin G. Zorn and Alexander Aiken, editors, *Proceedings of the 2010 ACM SIGPLAN Conference on Programming Language Design and Implementation, PLDI 2010, Toronto, Ontario, Canada, June 5-10, 2010*, pages 146–159. ACM, 2010. `doi:10.1145/1806596.1806615`.

**26**   Filip Pizlo, Lukasz Ziarek, and Jan Vitek. Real time Java on resource-constrained platforms with Fiji VM. In M. Teresa Higuera-Toledano and Martin Schoeberl, editors, *Proceedings of the 7th International Workshop on Java Technologies for Real-Time and Embedded Systems, JTRES 2009, Madrid, Spain, September 23-25, 2009*, ACM International Conference Proceeding Series, pages 110–119. ACM, 2009. `doi:10.1145/1620405.1620421`.

**27**   John H. Reppy. Concurrent ML: Design, Application and Semantics. In Peter E. Lauer, editor, *Functional Programming, Concurrency, Simulation and Automated Reasoning: International Lecture Series 1991-1992, McMaster University, Hamilton, Ontario, Canada*, volume 693 of *Lecture Notes in Computer Science*, pages 165–198. Springer, 1993. `doi:10.1007/3-540-56883-2\_10`.

**28**   Abhiroop Sarkar, Robert Krook, Bo Joel Svensson, and Mary Sheeran. Higher-Order Concurrency for Microcontrollers. In Herbert Kuchen and Jeremy Singer, editors, *MPLR '21: 18th ACM SIGPLAN International Conference on Managed Programming Languages and Runtimes, Münster, Germany, September 29-30, 2021*, pages 26–35. ACM, 2021. `doi:10.1145/3475738.3480716`.

**29**   Lui Sha, Ragunathan Rajkumar, and John P. Lehoczky. Priority Inheritance Protocols: An Approach to Real-Time Synchronization. *IEEE Trans. Computers*, 39(9):1175–1185, 1990. `doi:10.1109/12.57058`.

**30**   Bhargav Shivkumar, Jeffrey C. Murphy, and Lukasz Ziarek. RTMLton: An SML Runtime for Real-Time Systems. In Ekaterina Komendantskaya and Yanhong Annie Liu, editors, *Practical Aspects of Declarative Languages - 22nd International Symposium, PADL 2020, New Orleans, LA, USA, January 20-21, 2020, Proceedings*, volume 12007 of *Lecture Notes in Computer Science*, pages 113–130. Springer, 2020. `doi:10.1007/978-3-030-39197-3\_8`.

**31**   Vincent St-Amour and Marc Feeley. PICOBIT: A Compact Scheme System for Microcontrollers. In Marco T. Morazán and Sven-Bodo Scholz, editors, *Implementation and Application of Functional Languages - 21st International Symposium, IFL 2009, South Orange, NJ, USA, September 23-25, 2009, Revised Selected Papers*, volume 6041 of *Lecture Notes in Computer Science*, pages 1–17. Springer, 2009. `doi:10.1007/978-3-642-16478-1\_1`.

**32**   Mads Tofte and Jean-Pierre Talpin. Region-based Memory Management. *Inf. Comput.*, 132(2):109–176, 1997. `doi:10.1006/inco.1996.2613`.

**33**   Hideyuki Tokuda, Clifford W. Mercer, Yutaka Ishikawa, and Thomas E. Marchok. Priority Inversions in Real-Time Communication. In *Proceedings of the Real-Time Systems Symposium - 1989, Santa Monica, California, USA, December 1989*, pages 348–359. IEEE Computer Society, 1989. `doi:10.1109/REAL.1989.63587`.

**34**   Blase Ur, Elyse McManus, Melwyn Pak Yong Ho, and Michael L. Littman. Practical Trigger-Action Programming in the Smart Home. In Matt Jones, Philippe A. Palanque, Albrecht Schmidt, and Tovi Grossman, editors, *CHI Conference on Human Factors in Computing Systems, CHI'14, Toronto, ON, Canada - April 26 - May 01, 2014*, pages 803–812. ACM, 2014. `doi:10.1145/2556288.2557420`.

**35**   Steven Varoumas, Benoît Vaugon, and Emmanuel Chailloux. A Generic Virtual Machine Approach for Programming Microcontrollers: the OMicroB Project. In *9th European Congress on Embedded Real Time Software and Systems (ERTS 2018)*, 2018.

**36**   Benjamin Vedder, Jonny Vinter, and Magnus Jonsson. A Low-Cost Model Vehicle Testbed with Accurate Positioning for Autonomous Driving. *J. Robotics*, 2018:4907536:1–4907536:10, 2018. `doi:10.1155/2018/4907536`.

**37**   Ge Wang and Perry R. Cook. ChucK: A Concurrent, On-the-fly, Audio Programming Language. In *Proceedings of the 2003 International Computer Music Conference, ICMC 2003, Singapore, September 29 - October 4, 2003*. Michigan Publishing, 2003. URL: `http://hdl.handle.net/2027/spo.bbp2372.2003.055`.

**38**   Stephen Weeks. Whole-program compilation in MLton. In Andrew Kennedy and François Pottier, editors, *Proceedings of the ACM Workshop on ML, 2006, Portland, Oregon, USA, September 16, 2006*, page 1. ACM, 2006. `doi:10.1145/1159876.1159877`.

**39**   Gordon Williams. Espruino, 2012. URL: `http://www.espruino.com/`.

## A   Appendix A - Selective Communication and Events

To enable multi-party communication, synchronous message-passing models introduce a *selective communication* operator that races between two operations and selects the one that completes first. This enables the handling of communication with multiple participants without being unnecessarily blocked by the synchronous nature of the communication.

However, Reppy identified a conflict that arises between selective communication and procedural abstraction [27]. The complication can be demonstrated via the example of a client-server communication protocol where the client follows the protocol - *first send a message along a channel, `reqCh`, and only upon the success of the send will it accept the server response along the channel, `respCh`*. Such a protocol can be expressed in synchronous models like CSP and then abstracted as a procedure like the following:

```
clientCall : (Channel a, Channel b) -> a -> b
clientCall (reqCh, respCh) a =
  let _ = sendMsg reqCh a
   in recvMsg respCh
```

Now imagine a scenario where the client is communicating with two servers and the first server is temporarily unavailable. In such an instance the `sendMsg` call in line no. 3 will block and as the `sendMsg` call has been abstracted away inside the procedure it is not possible to apply the *select* operation on it. Hence, for liveness the `sendMsg` operation should not be hidden in the procedure.

However, if we expose the `sendMsg` operation, it goes against the principles of software abstraction where the internal operations of a protocol are leaked and can authorize the programmer to write unsafe operations like a sequence of two `sendMsg` calls that violates the protocol invariant (a send should always be followed by a receive). The `Event` construct of Concurrent ML resolves this issue elegantly by programming the abstraction as follows:

```
1 clientCallEvt : (Channel a, Channel b) -> a -> Event b
2 clientCallEvt (reqCh, respCh) a =
3   wrap (send reqCh a) (λ _ -> sync (recv respCh))
```

The `clientCallEvt` program above represents a server as a tuple of a request channel and a response channel. The use of `wrap` in `clientCallEvt` creates an event of type `Event b` (where `b` is the type of the values sent across the `respCh`). Send events have type `Event ()` so the function `(λ _ -> sync (recv respCh))` has type `() -> b`.

The `choose` operator, of type `choose : Event a -> Event a -> Event a`, can be used now. Given two servers, `server1 = (server1ReqCh, server1RespCh)` and `server2 = (server2ReqCh, server2RespCh)`, multi-party communication can be expressed without breaking the procedural abstraction using `choose (clientCallEvt server1) (clientCallEvt server2)`.

The return type of a `choose` call will still be `Event`, allowing us to compose and *choose* among several synchronous operations like *choose (choose (choose $ev_1$ $ev_2$) $ev_3$) $ev_4$*... When `sync` is applied to such a composite event it will race among all the events, $ev_1, ev_2, ev_3,..$, and synchronise on the operation that unblocks first.

## B  Appendix B - Scheduling Blinky

Fig. 14 below shows the timeline of our scheduler executing the blinky program from Listing 5. This chart involves two clocks. The actual wall-clock time, $T_{absolute}$, is represented along the X-axis while the process-local clock, $T_{local}$, for the process `foo` is shown inside the body of green chart representing `foo`.

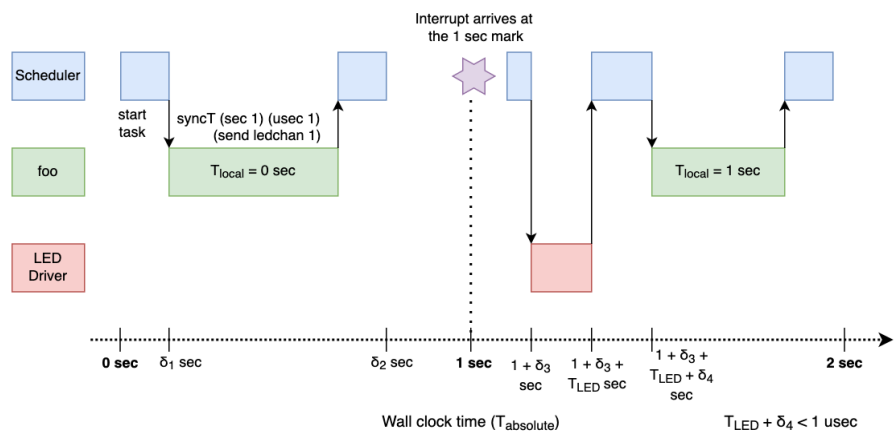

Figure 14  Scheduler timeline while executing the *blinky* program

When the program arrives at the `syncT` statement, an alarm is set for the time at which the VM should begin attempting communication with the LED driver. The alarm is set exactly at the 1-second mark, calculated from the $T_{local}$ clock, which removes the jitter associated with other statement executions and runtime overheads.

Once the alarm interrupt arrives, communication is initiated, and the deadline counter gets activated. The LED driver takes $T_{LED}$ seconds to execute, and the scheduler takes an additional $\delta_4$ time units to unblock the process `foo`. So, the deadline requested to the runtime follows the relation - $\delta_4 + T_{LED} < 1$ $usec$. Finally, $T_{local}$ is incremented by the relation $T_{local} = T_{local} + baseline$, where the baseline is 1 second in our case and the program continues.

## C  Appendix C - FSM Examples

### C.1  Four button blinky

**Listing 13** The complete Four-Button-Blinky program (Section 6.1) running on SynchronVM

```
butchan1 = channel ()
butchan2 = channel ()
butchan3 = channel ()
butchan4 = channel ()

ledchan1 = channel ()
ledchan2 = channel ()
ledchan3 = channel ()
ledchan4 = channel ()

press1 = wrap (recv butchan1) (λ x -> sync (send ledchan1 x))
press2 = wrap (recv butchan2) (λ x -> sync (send ledchan2 x))
press3 = wrap (recv butchan3) (λ x -> sync (send ledchan3 x))
press4 = wrap (recv butchan4) (λ x -> sync (send ledchan4 x))

anybutton = choose press1 (choose press2 (choose press3 press4))

program : ()
program =
  let _ = sync anybutton in
  program

main =
  let _ = spawnExternal butchan1 0 in
  let _ = spawnExternal butchan2 1 in
  let _ = spawnExternal butchan3 2 in
  let _ = spawnExternal butchan4 3 in
  let _ = spawnExternal ledchan1 4 in
  let _ = spawnExternal ledchan2 5 in
  let _ = spawnExternal ledchan3 6 in
  let _ = spawnExternal ledchan4 7 in
  program
```

### C.2  Large State Machine

**Listing 14** The complete complex state machine (Section 6.2) running on SynchronVM

```
butchan1 : Channel Int
butchan1 = channel ()
butchan2 : Channel Int
butchan2 = channel ()
butchan3 : Channel Int
butchan3 = channel ()
butchan4 : Channel Int
butchan4 = channel ()

ledchan1 : Channel Int
ledchan1 = channel ()
ledchan2 : Channel Int
```

```
13  ledchan2 = channel ()
14  ledchan3 : Channel Int
15  ledchan3 = channel ()
16  ledchan4 : Channel Int
17  ledchan4 = channel ()
18
19  not : Int -> Int
20  not 1 = 0
21  not 0 = 1
22
23  errorLed x = ledchan3
24
25  fail1ev = choose (wrap (recv butchan1) errorLed)
26           (choose (wrap (recv butchan3) errorLed)
27                   (wrap (recv butchan4) errorLed))
28
29  fail2ev = choose (wrap (recv butchan1) errorLed)
30           (choose (wrap (recv butchan2) errorLed)
31                   (wrap (recv butchan3) errorLed))
32
33  led1Handler x =
34     sync (choose (wrap (recv butchan2) (\x -> ledchan1)) fail1ev)
35
36  led2Handler x =
37     sync (choose (wrap (recv butchan4) (\x -> ledchan2)) fail2ev)
38
39  led : Int -> ()
40  led state =
41    let fsm1 = wrap (recv butchan1) led1Handler in
42    let fsm2 = wrap (recv butchan3) led2Handler in
43    let ch = sync (choose fsm1 fsm2) in
44    let _ = sync (send ch (not state)) in
45    led (not state)
46
47  main =
48    let _ = spawnExternal butchan1 0 in
49    let _ = spawnExternal butchan2 1 in
50    let _ = spawnExternal butchan3 2 in
51    let _ = spawnExternal butchan4 3 in
52    let _ = spawnExternal ledchan1 4 in
53    let _ = spawnExternal ledchan2 5 in
54    let _ = spawnExternal ledchan3 6 in
55    let _ = spawnExternal ledchan4 7 in
56    led 0
```

## D    Appendix D - The complete music programming example

We run this program on the STM32F4-discovery board that comes with a 12-bit digital-to-analog converter (DAC), which we connect to a speaker as a peripheral. We can write a value between 0 to 4095 to the DAC driver that gets translated to a voltage between 0 to 3V on the DAC output pin.

To produce a sound note we need to periodically write a sequence of 1's and 0's to the DAC driver. However, to make the produced note sound musical to the human ear, the *periodic rate* at which our process writes to the DAC driver is very important, and this is where the real-time aspect of the application comes in. The human ear recognises a note produced at a certain frequency as a musical note. Frequency is related to the periodic rate of a process by the relation:

$$Period = 1/Frequency$$

For instance, the musical note *A* occurs at a frequency of 440 Hz, which implies it has a time period of 2273 $\mu$seconds. From the point of view of the software, we are actually writing two values, a 1 and a 0, so we need to further divide the value by 2 to determine our rate of each individual write. If we call the rate of our writes as $Time_{Write}$, we get the relation -

$$Time_{Write} = Period/2 = 1/(2*Frequency)$$

Now that we know how to calculate the periodicity of our write in relation to the frequency, we need to know (i) what are the musical notes that occur in the "Twinkle, Twinkle" rhyme and (ii) what are the frequencies corresponding to those notes so that we can calculate the $Time_{Write}$ value from the frequency. The musical notes of the "Twinkle, Twinkle" tune (in the key of G) are well known and is given below:

**G G D D E E D C C B B A A G D D C C B B A D D C C B B A**

Given the above notes, the frequency of each of these notes are also well known. In Table 2 we show our calculation of the $Time_{Write}$ value for the various musical notes.

| Note | Frequency (Hz) | Period ($\mu$sec) | Time$_{Write}$ ($\mu$sec) |
| --- | --- | --- | --- |
| G | 196 | 5102 | 2551 |
| A | 220 | 4546 | 2273 |
| B | 247 | 4050 | 2025 |
| C | 261 | 3822 | 1911 |
| D | 294 | 3406 | 1703 |
| E | 329 | 3034 | 1517 |

**Table 2** Musical notes, their frequencies and time periods

Now we need to specify the time duration of each note. At the end of each note's duration period, we change the frequency of writes to the DAC driver. For instance, consider the transition from the second to the third note of the tune from G to D. If the note duration for G is 500 milliseconds then that implies our writing frequency should be 196 Hz for 500 milliseconds, and then at the 501st millisecond the frequency changes to 294 Hz (D's frequency).

When describing a musical etude, each note should be ideally mapped to its distinct duration in the program. A note duration can be a half note (1000 milliseconds) or a quarter note (500 milliseconds). The note duration of each of the 28 notes of the "Twinkle, Twinkle" tune is given below (Q implies a quarter note and H implies a half note):

**Q Q Q Q Q Q H Q Q Q Q Q Q H Q Q Q Q Q Q H Q Q Q Q Q Q H**

Listing 15 shows the entire program running on the SynchronVM that cyclically plays the "Twinkle, Twinkle, Little Stars" tune. The first 20 lines consists of declarations initialising a `List` data type and other standard library functions. Lines 54 - 68 consist of the principal logic of the program. Listing 15 can be compiled and run, **unaltered**, on an STM32F4-discovery board.

■ **Listing 15** The *Twinkle, Twinkle* tune (Section 6.3) running on SynchronVM

```
data List a where
  Nil  : List a
  Cons : a -> List a -> List a

head : List a -> a
head (Cons x xs) = x

tail : List a -> List a
tail Nil = Nil
tail (Cons x xs) = xs

not : Int -> Int
not 1 = 0
not 0 = 1

msec : Int -> Int
msec t = t * 1000

usec : Int -> Int
usec t = t

after : Int -> Event a -> a
after t ev = syncT t 0 ev

g  = usec 2551
a  = usec 2273
b  = usec 2025
c  = usec 1911
d  = usec 1703
e  = usec 1517
hn = msec 1000
qn = msec 500

twinkle : List Int
twinkle = Cons  g (Cons g (Cons d (Cons d (Cons e (Cons e (Cons d
          (Cons c (Cons c (Cons b (Cons b (Cons a (Cons a (Cons g
          (Cons d (Cons d (Cons c (Cons c (Cons b (Cons b (Cons a
          (Cons d (Cons d (Cons c (Cons c (Cons b (Cons b (Cons a Nil)
          ))))))))))))))))))))))))))))

durations : List Int
durations = Cons qn (Cons qn (Cons qn (Cons qn (Cons qn (Cons qn (Cons hn
             (Cons qn (Cons qn (Cons qn (Cons qn (Cons qn (Cons qn (Cons hn
             (Cons qn (Cons qn (Cons qn (Cons qn (Cons qn (Cons qn (Cons hn
             (Cons qn (Cons qn (Cons qn (Cons qn (Cons qn (Cons qn (Cons hn
             Nil))))))))))))))))))))))))))))

dacC : Channel Int
dacC = channel ()

noteC : Channel Int
noteC = channel ()

playerP : List Int -> List Int -> Int -> () -> ()
playerP melody nt n void =
  if (n == 29)
  then let _ = after (head nt) (send noteC (head twinkle)) in
       playerP (tail twinkle) durations 2 void
  else let _ = after (head nt) (send noteC (head melody)) in
       playerP (tail melody) (tail nt) (n + 1) void

tuneP : Int -> Int -> () -> ()
```

```
63  tuneP timePeriod vol void =
64    let newtp =
65        after timePeriod (choose (recv noteC)
66                                 (wrap (send dacC (vol * 4095))
67                                       (λ _ -> timePeriod))) in
68    tuneP newtp (not vol) void
69
70  main =
71    let _ = spawnExternal dacC 0 in
72    let _ = spawn (tuneP (head twinkle) 1) in
73    let _ = spawn (playerP (tail twinkle) durations 2) in
74    ()
```

Our application consists of two software processes and one external hardware process. The $Time_{Write}$ values of each of the twenty eight notes are represented as the list `twinkle` on Lines 34-39 and the note durations are contained in the `durations` list (Lines 41-46). We use two channels - `dacC` to communicate with the DAC and `noteC` to communicate between the two software processes.

Owing to the different time periods of the two processes, their $T_{local}$ clock progresses at different rates. In Figure 15 we visualise the message passing that occurs between the two software process and the hardware process when transitioning from a note C4 to a note G4.

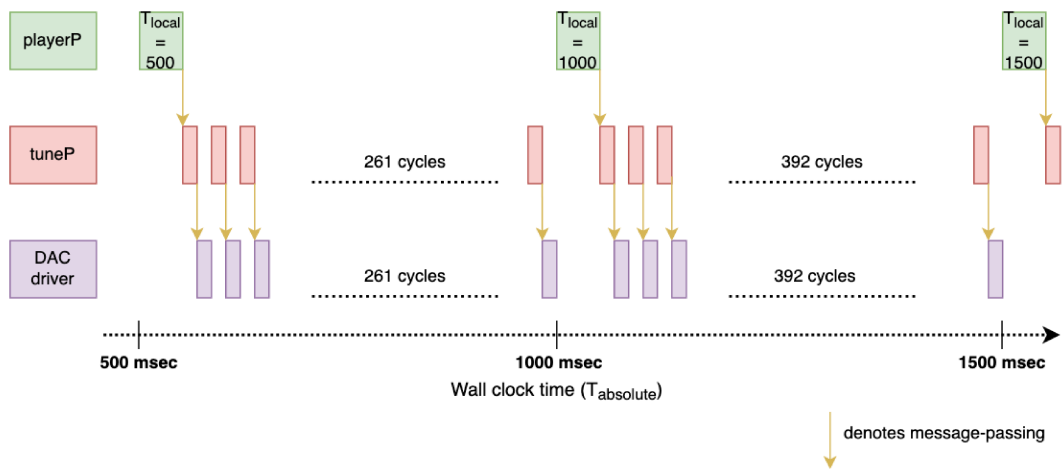

**Figure 15** Moving from the note C4 to note G4

As the `playerP` process runs once every 500 milliseconds, the `tuneP` process completes $500 * 10^3/1915 = 261 cycles$ when playing the note C. For the next note, G, the $Time_{Write}$ value changes to 1432 microseconds and the corresponding write frequency changes to 392 cycles and the process cyclically carries on.